\documentclass[11pt]{article} 
\usepackage{a4wide,epsf,cite} 
\usepackage[tbtags]{amsmath}

\newcommand{\tr}{{\rm{tr}}}  
 
\newcommand{\g}{g}

\begin{document}

\hbox{}  
\nopagebreak  
\vspace{-3cm} 
\begin{flushright} 
  { OUTP-00 10P} \\  
  { NORDITA-2000/14 HE} 
\end{flushright} 
 
\vspace{1in} 
 
\begin{center} 
  {\Large \bf Relating different approaches to nonlinear QCD evolution
    at finite gluon density} \vspace{0.1in}
   
  \vspace{0.5in} {\large Alex Kovner$^{1)}$, J. Guilherme
    Milhano$^{1)}$
    and Heribert Weigert$^{2)}$}\\
  {\small $^{1)}$ Theoretical Physics, Oxford University, 1 Keble
    Road, Oxford,
    OX1 3NP, UK\\
    $^{2)}$ NORDITA,  Blegdamsvej 17, 2100 Copenhagen \O, Denmark  }\\
  \vspace{0.5in}

\vspace{.2in}

\begin{minipage}{.9\textwidth} 
  {\small We analyze the relation between evolution equations at low 
    $x$ that have been derived in different approaches in the last 
    several years.  We show that the equation derived by Balitsky and 
    Kovchegov is obtained from the 
    Jalilian-Marian--Kovner--Leonidov--Weigert (JKLW) equation in the 
    limit of small induced charge density.  We argue that the higher 
    nonlinearities resummed by the JKLW equation correspond, in 
    physical terms, to the breakdown of the eikonal approximation when 
    the gluon fields in the target are large.} 
\end{minipage} 
\end{center}

\section{Introduction} 
\label{sect:intro}

In recent years, there has been renewed interest in the understanding 
of the physics of systems with large number of partons.  These studies 
have been essentially motivated by two large experimental programmes 
--- low $x$ deep inelastic scattering (DIS) at HERA and heavy ion 
collision experiments at RHIC and LHC.  Both physical situations 
involve a large number of participating gluons.  In low $x$ DIS these 
gluons are generated in the proton light cone wave function by the 
evolution to low $x$, whereas in the nuclear collision this evolution 
is enhanced since the nuclear wave function contains many gluons 
already at moderate values of energy. 
 
The growth of gluon density leads to interesting physical
consequences, the physical understanding of which has been steadily
improving.  One universal feature now believed to be true, is the
saturation of gluon densities.  Apparently, the number of gluons per
unit phase space volume practically saturates and at large densities
grows only very slowly (logarithmically) as a function of the
parameter that triggers the growth. The relevant parameter could be
$1/x$ in the low $x$ regime or the atomic number of the nucleus $A$ in
heavy ion collisions. This saturation takes place at values of
transverse momentum below a certain saturation momentum $k_s$, which
itself depends on $1/x$ and $A$.  The nature of this dependence is
less well understood. In the analysis based on the
Balitsky-Fadin-Kuraev-Lipatov (BFKL) evolution \cite{Mueller:1999wm}
and on the double logarithmic approximation (DLA) \cite{moriond} the
dependence is powerlike $k_s\propto(1/x)^{\alpha_s\delta}$, while
other approaches \cite{Capella:1999kv} suggest a much slower dependence.  In
the case of power dependence, the saturation momentum at HERA is
estimated to be in the range of 1-2 GeV with similar, slightly higher,
values at LHC.  Optimistically, one can hope that the saturation
region is itself semiperturbative, that is the value of the coupling
constant is reasonably small and, therefore, weak coupling methods can
be applied to the quantitative analysis of the phenomenon.
 
The physics of saturation must have experimental manifestations. 
The simplest and the most direct, in a way, is the unitarization of 
the total DIS cross section.  This is, however, also the least 
interesting one. First, since the effect of unitarization is almost 
kinematical, one does not need high partonic density, it is enough to 
have a large number of partons, not necessarily in the same bin of the 
phase volume \cite{Mueller:1995gb,Kovchegov:1997dm}. Second, because the experimental 
status of unitarization is unclear. So far, all DIS data on the total 
cross section can be reasonably well described by linear 
Dokshitzer-Gribov-Lipatov-Altarelli-Parisi (DGLAP) evolution without 
the need to include nonlinear effects \cite{Martin:1998sq}. Although physically 
it is hard to believe that the leading twist perturbative approximation 
can be applied at $Q^2$ as low as 1 GeV$^2$ and although some aspects 
of the gluon distribution that emerge form these fits \cite{Martin:1998sq} are 
intuitively not satisfactory, present inclusive DIS data cannot be 
considered as an unambiguous confirmation of nonlinear effects. 
 
The realm of nonlinear effects is, however, much richer than the total 
cross section.  In particular, one expects qualitative changes in the 
structure of the final states as one moves into the saturation region. 
The study of these effects has, however, not started in earnest yet 
and we have a long way to go before being able to make verifiable 
quantitative predictions. 
 
In particular, one needs a well defined formal framework to perform 
calculations.  Several approaches to the problem have been developed 
in recent years by different groups. They all rely on the smallness of 
the coupling constant while resumming the effects of large number of 
partons/partonic density.  The aim of all these approaches is 
essentially to derive the evolution of the hadronic scattering cross 
section with $1/x$.  They, however, utilize different techniques and 
conceptual frameworks and the resulting evolution equations look 
rather different.  It is the aim of this paper to explore the relation 
between some of these different approaches in an attempt to understand 
where they diverge from each other in terms of physics input. 
 
In particular, we will concern ourselves with three recent works 
Ref.~\cite{Balitskii:1996ub}, Ref.~\cite{Kovchegov:1999yj} and 
Refs.~\cite{Jalilian-Marian:1997jx,Jalilian-Marian:1997gr,Jalilian-Marian:1997dw,Kovner:1999bj,Jalilian-Marian:1998cb}.  In 
Ref.~\cite{Balitskii:1996ub} the evolution equation for the scattering 
amplitude is derived using the effective action and the eikonal 
approximation in the target rest frame. Ref.~\cite{Kovchegov:1999yj} uses the 
dipole model method of Refs.~\cite{mueller:1994rr,Mueller:1994jq}.  And, finally, 
Refs.~\cite{Jalilian-Marian:1997jx,Jalilian-Marian:1997gr,Jalilian-Marian:1997dw,Kovner:1999bj} uses the effective action 
in the projectile rest frame to derive the evolution of the hadron 
light cone wave function with $1/x$.  We will refer to the resulting 
evolution equation as the Jalilian-Marian--Kovner--Leonidov--Weigert 
(JKLW) equation. 
 
The outline of this paper is the following. In Sec.~\ref{sect:bk} we 
rederive the evolution equation of Ref.~\cite{Balitskii:1996ub} in a simple 
and intuitive way.  This derivation makes it obvious that this 
approach is equivalent to the approach of Ref.~\cite{Kovchegov:1999yj} up to 
subleading corrections in $1/N_c$. This is not new and was noted 
already in Ref.~\cite{Kovchegov:1999yj}. In the following, we will refer to this 
evolution equation as the Balitsky-Kovchegov (BK) equation\footnote{The 
$O(1/N_c)$ differences between the equations derived in Ref.\cite{Kovchegov:1999yj} and
\cite{Balitskii:1996ub} do not carry essential new physics. They therefore
do not affect our understanding of the relationship between the generic frameworks of 
the BK and JKLW equations.}.
We discuss the physical picture of this evolution and resulting 
unitarization of the total cross section in both target and projectile 
rest frames and point out the effects due to which the approximations 
involved should break down at extremely small $x$.  The breakdown of 
the approximation should have very little effect on the unitarization 
of the total cross section, since, especially for large targets like 
nuclei, the black disk limit should be reached while the approximation 
is still valid. However, one does expect the structure of the final 
states to be strongly affected. Our discussion here is, in large 
measure, parallel to that of Refs.~\cite{Mueller:1995gb,Kovchegov:1997dm}. 
 
In Sec. \ref{sect:jklw} we relate explicitly the calculation of 
Ref.~\cite{Balitskii:1996ub} to that of 
Refs.~\cite{Jalilian-Marian:1997jx,Jalilian-Marian:1997gr,Jalilian-Marian:1997dw,Kovner:1999bj}. In particular, we 
calculate the basic physical quantities appearing in the evolution 
equation of Refs.~\cite{Jalilian-Marian:1997jx,Jalilian-Marian:1997gr,Jalilian-Marian:1997dw,Kovner:1999bj} in the 
approximation of Ref.~\cite{Balitskii:1996ub}. We show that the results of 
Ref.~\cite{Balitskii:1996ub} are recovered from 
Refs.~\cite{Jalilian-Marian:1997jx,Jalilian-Marian:1997gr,Jalilian-Marian:1997dw,Kovner:1999bj} in the limit of small 
induced fields. We also show that the double logarithmic limit of the 
evolution of Ref.~\cite{Balitskii:1996ub} is trivial. That is, in the double 
logarithmic limit, the evolution equation for the gluon distribution 
function (defined operatorially as the number of gluons in the 
lightcone gauge in the infinite momentum frame) becomes linear and 
does not contain any Gribov-Levin-Ryskin (GLR) type 
corrections\footnote{This is not to say that the evolution of the DIS 
  cross section for which the equation of Ref.~\cite{Balitskii:1996ub} and 
  \cite{Kovchegov:1999yj} has been derived is linear in the double logarithmic 
  limit.  The GLR type nonlinearity does indeed appear in the 
  evolution equation for the virtual photon cross section due to the 
  nonlinear relation between the cross section and the gluon 
  distribution function.}. This is in contrast with the result of 
Ref.~\cite{Jalilian-Marian:1998cb} where it was shown that the double logarithmic limit 
of the evolution Refs.~\cite{Jalilian-Marian:1997jx,Jalilian-Marian:1997gr,Jalilian-Marian:1997dw,Kovner:1999bj} results 
in a nonlinear equation.  We point out that this is indeed a very 
natural result from the point of view of the dipole model approach. 
 
In Sec.~\ref{sect:more} we transform the full calculation of 
Refs.~\cite{Jalilian-Marian:1997jx,Jalilian-Marian:1997gr,Jalilian-Marian:1997dw,Kovner:1999bj} into the framework of 
Ref.~\cite{Balitskii:1996ub}. We show that in the approach of 
Ref.~\cite{Balitskii:1996ub} it corresponds to abandoning the eikonal 
approximation or, equivalently, to the inability to fully describe the 
target by a classical $A^+$ field.  We also point out technical 
reasons which lead us to believe that, in fact, in the framework of 
the effective action of Ref.~\cite{Balitskii:1996ub} such a failure is 
expected when the evolution is continued to very low values of $x$. 
 
Finally, we conclude with a brief discussion in Sec. \ref{sect:conc}.

\section{A simple derivation of the BK equation} 
\label{sect:bk} 
 
In this section we will give a simple derivation of the evolution 
equation first derived in Ref.~\cite{Balitskii:1996ub} and discuss the 
physical picture behind it.  Consider the deep inelastic scattering at 
low $x$.  We will work in the frame in which the photon fluctuates 
into an energetic quark-antiquark pair long before it reaches the 
target, but where most of the energy resides in the target hadron 
which moves very fast.  The scattering of the quark-antiquark pair is 
dominated by its interaction with the gluons in the target.  Since the 
target hadron moves fast, the time evolution of the gluon fields is 
slowed by Lorentz time dilation.  Also, due to Lorentz contraction, 
the gluon fields are well localized in the plane perpendicular to the 
direction of motion, which we take to be the positive $x_3$ axis.  The 
target can, therefore, be modeled by a distribution of static gluon 
fields localized at $x^-=0$.  As the scattering energy increases ($x$ 
decreases) the gluon fields of the target change due to contributions 
of quantum fluctuations.  It is this evolution in $x$ of the hadronic 
ensemble that we intend to describe in terms of the evolution 
equation. 
 
\subsection{The BK equation.}

In this section we will use the lightcone gauge $A^-=0$.  In this
gauge, following Ref.~\cite{Balitskii:1996ub}, we take the vector potentials
representing the relevant gluon field configurations to be of the form
\begin{equation} 
\label{aplus} 
  b^i=0,\qquad b^+=b(x_\perp)\delta(x^-)\, . 
\end{equation} 
Here and in the rest of this section, unless otherwise specified, we
use the matrix notation for the gauge field $b^+=b^+_a t_a$ etc.,
where $t_a$ are the generators of the $SU(N)$ group in the fundamental
representation.  One can reasonably ask whether the vector potential
of this form is the only relevant one.  This turns out to be a
nontrivial question. In fact, we will argue later in
Sec.~\ref{sect:more} that this is not quite the case if we want to be
able to describe the evolution up to arbitrarily small values of $x$.
At this point, however, we follow Ref.~\cite{Balitskii:1996ub}.  We will
return to this question in Sec.~\ref{sect:more}.
 
The DIS structure function can be written in the following general
form
\begin{equation} 
  F_2(x,Q^2)=\frac{Q^2}{ 4\pi^2\alpha_{em}}  
  \int\frac{dzdx_\perp dy_\perp}{ 4\pi} 
  \Phi(x_\perp-y_\perp,z)N(x_\perp,y_\perp,y)\, . 
\end{equation} 
Here, $x_\perp$ and $y_\perp$ are the transverse coordinates of the
quark and the antiquark in the pair, $z$ is the fraction of the pairs
longitudinal momentum carried by the quark and $y$ is the rapidity of
the slowest particle in the pair. Also, $\Phi(x_\perp-y_\perp,z)$ is
the square of the ``wave function'' of the photon --- the probability
that the virtual photon fluctuates into the pair with given
coordinates and momenta --- and $N(x_\perp,y_\perp,y)$ is the cross
section for the scattering of the pair.
 
The wave function $\Phi$ is well known.  It is given for example in
Ref.~\cite{Kovchegov:1999yj}, but its explicit form will not be of interest to us.
We concentrate our discussion on the scattering cross section $N$.  If
the quark-antiquark pair is energetic enough the scattering cross
section is eikonal
\begin{equation}  
  N(x_\perp,y_\perp)=\tr \big\langle  
  V(x_\perp ) V^\dagger (y_\perp )-1\big\rangle_A \, , 
\label{n} 
\end{equation} 
where $V$ ($V^\dagger$) is the eikonal phase for the scattering of the
energetic quark (antiquark)
\begin{equation}  
  V(x^+=0,x_\perp ) = {\cal P}  
  \exp \Big[ -i \g\int\limits_{-\infty}^{+\infty}\!\! dx^-  
  A^+ (x^+=0, x_\perp , x^-) \Big] 
\label{v} 
\end{equation} 
with the vector potential in the fundamental representation.  We,
therefore, have to calculate the average of $\big\langle V(x^+ = 0,
x_\perp) V^\dagger(y^+ = 0, y_\perp)\big\rangle_A$ over the hadronic
wave function as indicated by $\langle\ldots\rangle_A$. In our frame,
the quark and the antiquark move with velocity of light in the
negative $x_3$ direction. All the fields in Eq.~(\ref{v}), therefore,
have vanishing $x^+$ coordinate. This will also be the case for all
the fields in the rest of this section. For simplicity, we suppress
the $x^+$ coordinate in the following.
 
In the leading approximation the vector potential is given by
Eq.~(\ref{aplus}) and the scattering amplitude is
\begin{equation} 
\label{Nb} 
N(x_\perp,y_\perp)=\big\langle  
\tr [ U(x_\perp ) U^\dagger (y_\perp ) -1]\big\rangle_b 
\end{equation} 
with
\begin{equation} 
  U(x_\perp ) = {\cal P} \exp  
\Big[ - i \int\limits_{-\infty}^{+\infty}\!\! dx^-  
b^+ (x_\perp , x^-) \Big]\, , 
\label{uf} 
\end{equation} 
To calculate the order $\alpha_s$ correction to this expression we
write the vector potential as
\begin{equation} 
  A^+=\frac{1}{g} b^+ + a^+ 
\end{equation} 
with $a^+$ being a small fluctuation and expand the eikonal factors to
second order in $a^+$.

Recalling that the classical background vector potential is a delta
function in $x^-$, we have
\begin{subequations} 
\begin{align}\label{vexp} 
  V(x_\perp ) = &\, {\cal P}\exp \Big[ - i\g
  \int\limits_{-\infty}^{0}\!\! dx^- a^+ (x_\perp , x^-) \Big]
  U(x_\perp ) {\cal P} \exp \Big[ - i\g
  \int\limits_{0}^{+\infty}\!\! dx^- a^+ (x_\perp , x^-) \Big] \\
  = & \, U(x_\perp ) - i\g \Big\{ \int\limits_{-\infty}^{0}\!\! dx^- a^+
  (x_\perp , x^-) U(x_\perp ) + U(x_\perp )
  \int\limits_{0}^{+\infty}\!\! dx^-
  a^+ (x_\perp , x^-) \Big\} \nonumber \\
  & \phantom{U(x_\perp )\:} - \g^2\Big\{ \int\limits_{-\infty}^{0}\!\!
  dx^- dy^- \theta (y^- - x^-)
  a^+ (x_\perp , x^-) a^+ (x_\perp , y^-) U(x_\perp ) \nonumber \\
  & \phantom{U(x_\perp )\:-\Big\{} + \int\limits_{-\infty}^{0}\!\!
  dx^- a^+ (x_\perp , x^-) U(x_\perp )
  \int\limits_{0}^{+\infty}\!\! dx^- a^+ (x_\perp , x^-) \nonumber \\
  & \phantom{U(x_\perp )\:-\Big\{} + U(x_\perp
  )\int\limits_{0}^{+\infty}\!\! dx^- dy^- \theta (y^- - x^-) a^+
  (x_\perp , x^-) a^+ (x_\perp , y^-) \Big\}\, .
\end{align} 
\end{subequations} 
All contributions break down into $x^-$ ordered pieces because of the
$x^-$ structure in Eq.~(\ref{aplus}).
 
Now, together with the analogous expansion for $V^\dagger$, we insert
this into Eq.~(\ref{n}) and obtain
\begin{align} 
  \tr \big\langle V(x_\perp ) V^\dagger (y_\perp &) \big\rangle_A
  -\tr \big\langle U(x_\perp ) U^\dagger (y_\perp ) \big\rangle_b = \nonumber \\
  = \g^2\tr\Big\langle & \int_{-\infty}^{0} \!\! dw^- a^+ (x_\perp , w^-)
  U(x_\perp )
  U^\dagger (y_\perp ) \int_{-\infty}^{0} \!\! dz^- a^+ (y_\perp , z^-) \nonumber \\
  & + \int_{-\infty}^{0}\!\! dw^- a^+ (x_\perp , w^-) U(x_\perp )
  \int_{0}^{+\infty}\!\! dz^-
  a^+ (y_\perp , z^-) U^\dagger (y_\perp ) \nonumber \\
  & + U(x_\perp ) \int_{0}^{+\infty}\!\! dw^- a^+(x_\perp , w^-)
  U^\dagger ( y_\perp )
  \int_{-\infty}^{0}\!\! dz^- a^+ (y_\perp , z^-) \nonumber \nonumber \\
  & + U(x_\perp ) \int_{0}^{+\infty} \!\! dw^- a^+ (x_\perp , w^-)
  \int_{0}^{+\infty} \!\! dz^- a^+ ( y_\perp , z^-) U^\dagger (y_\perp )  \nonumber\\
  & - \int_{-\infty}^{0} \!\! dw^- dz^- \theta (z^- - w^-) a^+
  (x_\perp , w^-) a^+ (x_\perp , z^-)
  U( x_\perp )  U^\dagger (y_\perp ) \nonumber \\
  & - \int_{-\infty}^{0}\!\! dw^- a^+ (x_\perp , w^-) U(x_\perp )
  \int_{0}^{+\infty}\!\!  dz^-
  a^+ (x_\perp , z^-) U^\dagger (y_\perp ) \nonumber \\
  & - U( x_\perp ) \int_{0}^{+\infty} \!\! dw^- dz^- \theta (z^- -
  w^-) a^+ (x_\perp ,
  w^-) a^+ (x_\perp , z^-) U^\dagger (y_\perp ) \nonumber \\
  & - U( x_\perp ) U^\dagger ( y_\perp ) \int_{-\infty}^{0} \!\! dw^-
  dz^- \theta (w^- - z^-)
  a^+(y_\perp , w^-) a^+ (y_\perp , z^-)  \nonumber \\
  & - U(x_\perp ) \int_{0}^{+\infty} \!\! dw^- a^+ ( y_\perp , w^-)
  U^\dagger (y_\perp )  \int_{-\infty}^{0} \!\! dz^- a^+ (y_\perp , z^-)  \nonumber \\
  & - U(x_\perp ) \int_{0}^{+\infty} \!\! dw^- dz^- \theta (w^- - z^-)
  a^+ ( y_\perp , w^-) a^+ ( y_\perp , z^-) U^\dagger (y_\perp )
  \Big\rangle_{b,a}.
\label{expansion} 
\end{align} 
 
In writing Eq.~(\ref{expansion}) we have anticipated that $\big\langle
a^+_u\big\rangle_a=0$ as in the free case. This can easily be shown
using the explicit expression for the fluctuation propagator given
below.

Although this expression is a little cumbersome, the physical meaning
of the various terms is very clear.  Put more compactly the structure
of the above is determined by
\begin{equation}  
\label{revirt} 
\begin{split} 
  \tr \big\langle V(x_\perp ) V^\dagger (y_\perp &) \big\rangle_A -\tr
  \big\langle U(x_\perp ) U^\dagger (y_\perp ) \big\rangle_b = \g^2
  \Big\langle\ \ \big\langle a^+_u a^+_v\big\rangle_a\ \ 
  \frac{1}{2}\Big( 2\frac{\delta}{\delta b^+_u}
  U(x_\perp)\frac{\delta}{\delta b^+_v} U^\dagger(y_\perp) \\ &
  +\big(\frac{\delta}{\delta b^+_u} \frac{\delta}{\delta b^+_v}
  U(x_\perp)\big) U^\dagger(y_\perp) + U(x_\perp)
  \big(\frac{\delta}{\delta b^+_u} \frac{\delta}{\delta b^+_v}
  U^\dagger(y_\perp)\big) \Big)\ \Big\rangle_b\, .
\end{split} 
\end{equation} 
Diagrammatically the r.h.s. can be represented as follows
\begin{equation} 
\label{revirt_diags} 
  \big\langle\ \ 2\times 
  \begin{minipage}[m]{3cm}\epsfxsize=3cm  
    \epsfbox{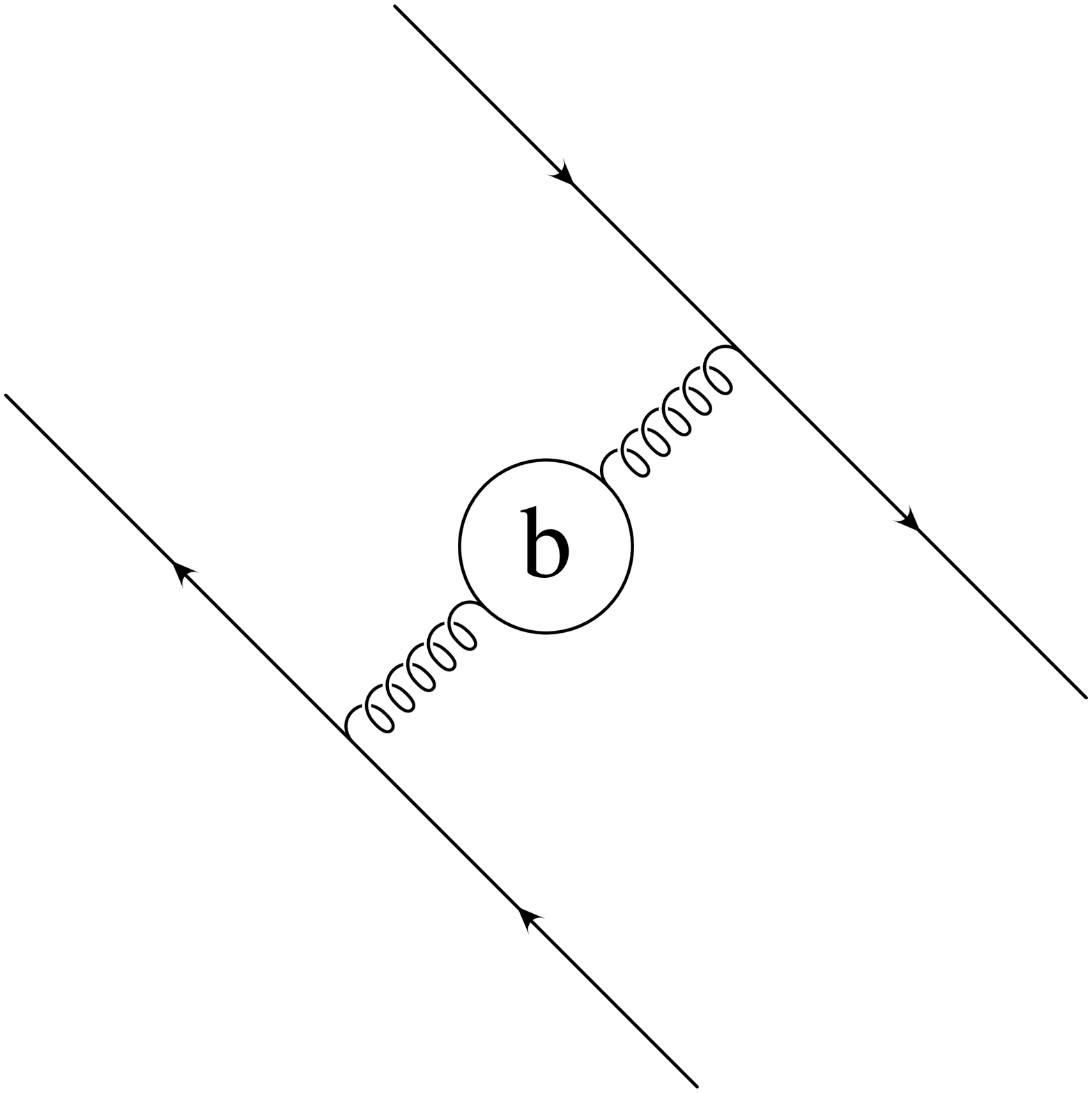}
  \end{minipage} 
  + 
  \begin{minipage}[m]{3cm}\epsfxsize=3cm  
    \epsfbox{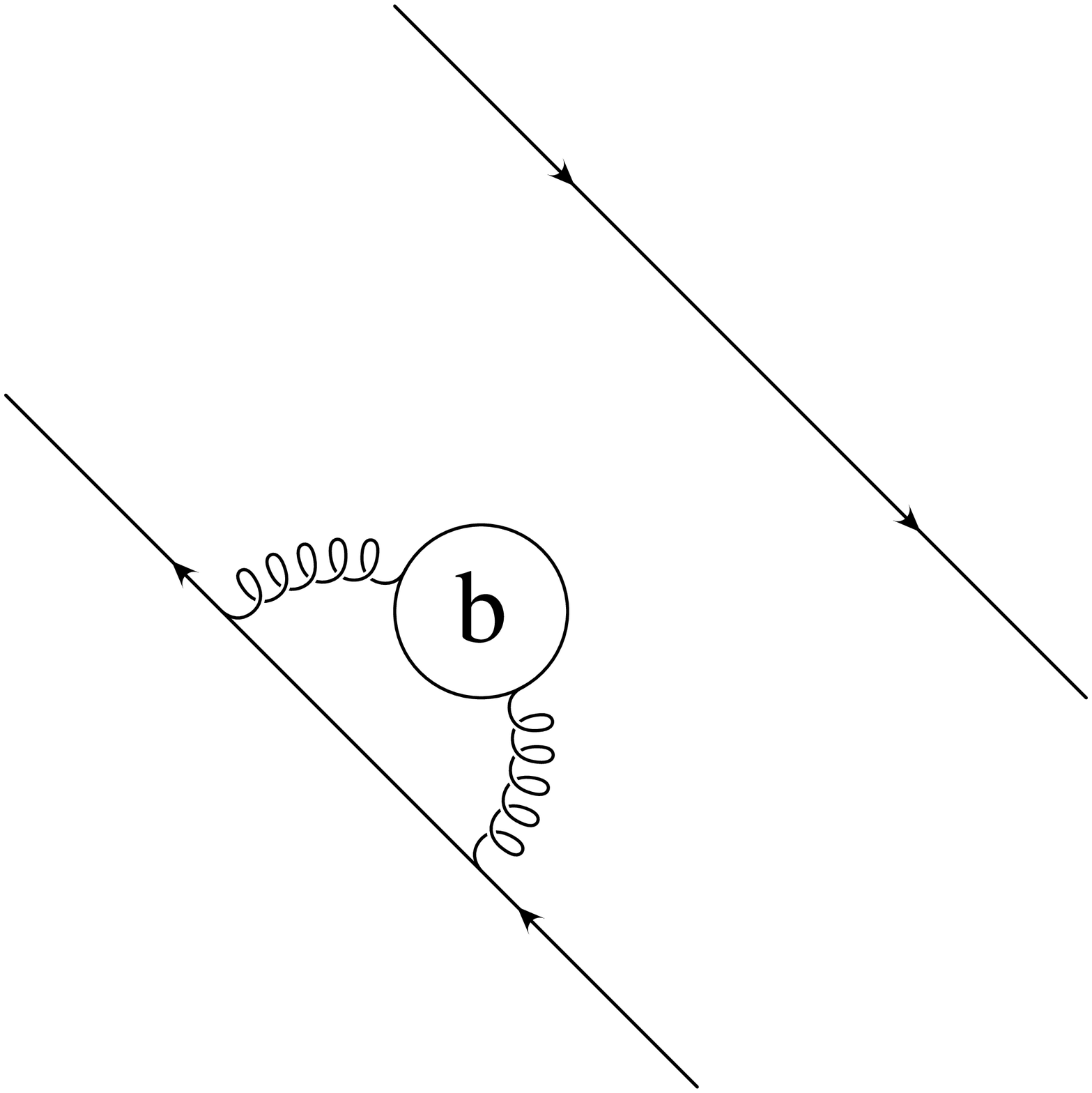}
  \end{minipage} 
  +\begin{minipage}[m]{3cm}\epsfxsize=3cm  
    \epsfbox{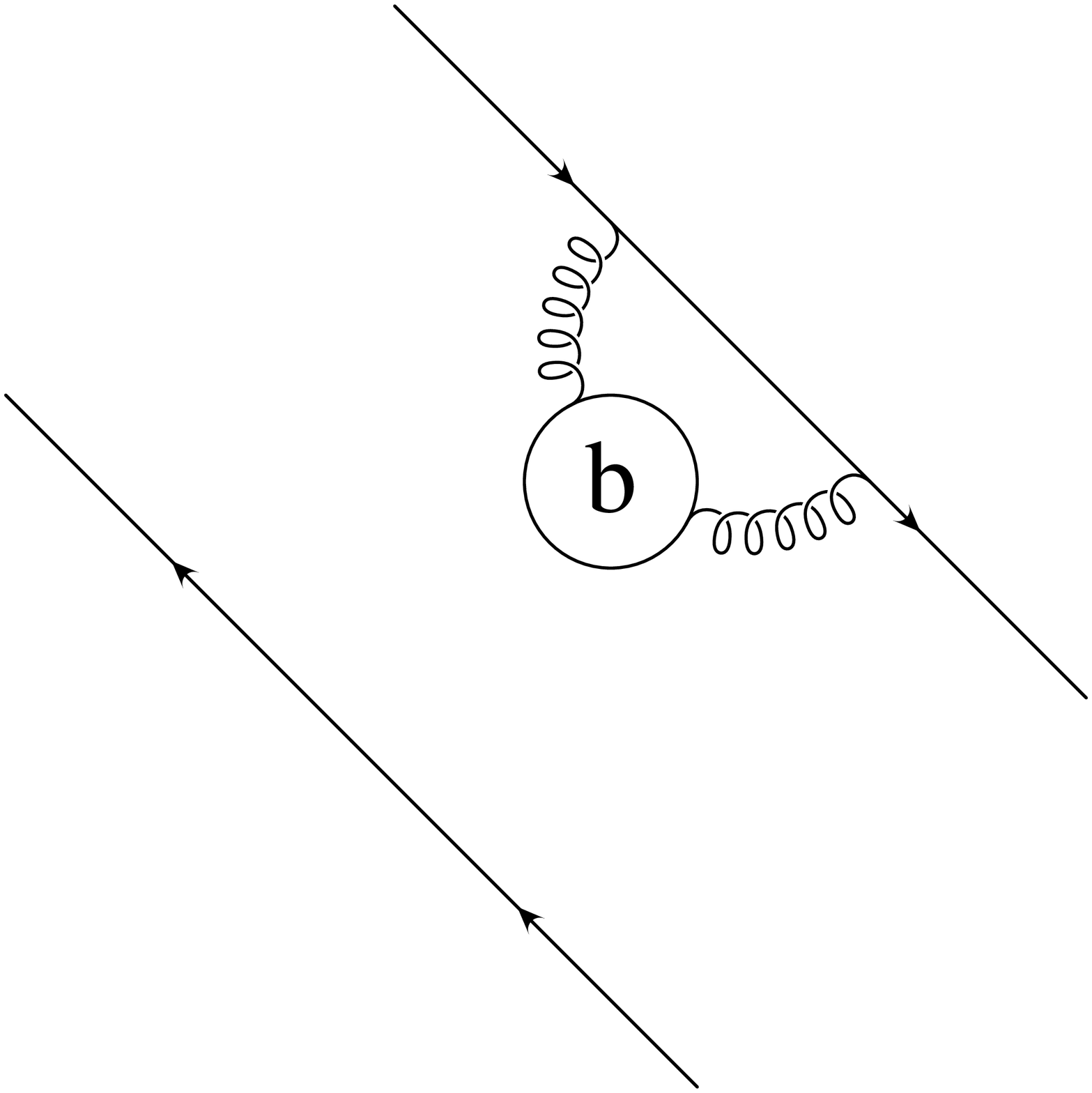} 
  \end{minipage}\ \ \big\rangle_b 
\end{equation} 
The straight lines represent the eikonal factors $U$, while the curly
lines denote the gluon fluctuation propagator $\big\langle a^+_u
a^+_v\big\rangle_a$, evaluated in the fixed background $b^+(x)$.  The
terms in Eq.~(\ref{revirt}) with first order derivatives correspond to
processes where the gluon is emitted by the quark and absorbed by the
antiquark (or vice versa). Those will be hereafter referred to as
``exchange contributions''.

The terms in Eq.~(\ref{revirt}) with second derivatives acting on $U$
(or $U^\dagger$) correspond to the diagramms where where the quark (or
antiquark) emits the gluon and then reabsorbs it at a later time;
typical self energy corrections. To contrast them against the exchange
contributions we will also refer to them as ``non-exchange contributions''.

In Eq.(\ref{expansion}) we are looking at a $x^-$ ordered breakdown of
the diagrams in Eq.~(\ref{revirt_diags}) with the first four terms
summing up to the exchange contribution and the rest to the
non-exchange contributions.
 
With the vertices known, we only lack an explicit expression for
$\big\langle a^+_u a^+_v\big\rangle_a$.  The QCD action expanded to
second order in the fluctuation field $a^+$, in the presence of the
classical background $b^+$, in our lightcone gauge is
\begin{equation} 
  \begin{split} 
    S = \frac{1}{2} \bigg\{ & a^{+}_{a} \big[ - (
    {\partial}^{-})^{2} \big] a^{+}_{a}
    - 2 ({\partial}^{i} a^{+}_{a})({\partial}^{-} a^{i}_{a}) \\
    & - a^{i}_{a} \Big[ \big( 2 D^{+}_{ab}[b]{\partial}^{-} -
    ({\partial}_{\perp})^{2} {\delta}_{ab} \big) {\delta}^{ij} +
    {\partial}^{i}{\partial}^{j} {\delta}_{ab} \Big] a^{j}_{b}
    \bigg\}\, .
\end{split} 
\label{action} 
\end{equation} 
Recall that we are interested only in the propagator of the fields at
equal $x^+$. Consequently, it is only the on-shell part of the
propagator that is relevant for our purposes. We can, therefore, use
the classical equation of motion for $a^{+}$
\begin{equation} 
  a^{+} = \frac{\partial^i}{\partial^-} a^i\, . 
\label{a+} 
\end{equation} 
Substituting this in Eq.~(\ref{action}) we get
\begin{equation} 
  S = - \frac{1}{2} a^{i}_{a} 
  (D^{2})_{ab} {\delta}^{ij} a^{j}_{b}\, , 
\end{equation} 
where
\begin{equation} 
  (D^{2})_{ab} = 2 D^{+}_{ab}[b] {\partial}^{-}  
  - ({\partial}_{\perp})^{2} {\delta}_{ab} \, . 
\end{equation} 
The propagator of the spatial components of the vector potential is,
therefore, simply given by $-i/D^2$. The explicit form is very simple
and can be found, for example, in Ref.~\cite{Hebecker:1998kv}.
\begin{equation} 
  \begin{split} 
    -i \left[\frac{1}{(D^{2})}\right]_{ab} = \int\frac{dp^-}{2 p^-
      (2\pi)^3} & \big[\theta (x^- - y^-) \theta (p^-) - \theta (y^- -
    x^-)\theta (-p^-) \big] \times
    \\
    \times \int d^2 & p_{\perp} d^2 q_{\perp} e^{-ip\cdot x+iq\cdot y}
    \int \frac{d^2 z_{\perp}}{(2\pi)^2} e^{-i( p_{\perp} - q_{\perp})
      z_{\perp}} \tilde{U}^{-1}_{ab} (x^-, y^-, z_{\perp})
  \end{split} 
\end{equation} 
with ${p}^{+} =\frac{{p}_{\perp}^{2}}{2p^-}$, ${q}^{+}
=\frac{{q}_{\perp}^{2}}{2p^-}$ and ${q}^{-} = {p}^{-}$. The adjoint
color matrix\footnote{More rigorously, the structure of $
  \tilde{U}^{-1}$ is given by $\tilde U_{ab}^{-1}(z_\perp) =
  e^{i(\theta(x^-) - \theta(y^-)) b(z_\perp)}$.  However, the
  difference between this expression and that given in
  Eq.(\ref{tUrep}) only shows up if it is multiplied by $\partial^+$
  derivatives or $\delta(x^-)$ factors. Since we encounter no such
  factors in our calculation we will be using Eq.(\ref{tUrep})
  throughout.}
\begin{equation} 
\label{tUrep} 
  \begin{split} 
    \tilde{U}^{-1}_{ab} (x^-, y^-, z_{\perp}) =
    \big(\theta(x^-)\theta(&y^-)+\theta(- x^-)\theta(-y^-)\big)\delta_{ab} \\
    &+\theta(-x^-)\theta(y^-) \tilde{U}_{ab} (z_{\perp})
    +\theta(x^-)\theta(-y^-) \tilde{U}^{\dagger}_{ab}(z_{\perp})\, .
  \end{split} 
\end{equation} 
represents a phase factor one picks up when crossing the $x^-=0$ plane
due to interaction with a field of type Eq.~(\ref{aplus}).  Here,
$\tilde{U}_{ab}(z_{\perp})$ is the adjoint version of the fundamental
$U$ in Eq.~(\ref{uf}).  If the $x^-=0$ plane is not crossed the
propagation remains free.
  
We can now write the on-shell correlator of the ``$+$'' component of
the vector potential as
\begin{multline} 
  \big\langle a_a^+ (x^+ = 0, x_\perp , x^-)
  a_b^+ (y^+ = 0, y_\perp , y^-)\big\rangle_a\\
  = \big\langle \frac{\partial^i_x}{\partial^-_x} a_a^i (x^+ = 0,
  x_\perp , x^-) a_b^i (y^+ = 0, y_\perp , y^-)
  \frac{\partial^i_y}{\partial^-_y}
  \big\rangle  \\
  = - \partial^i_x \partial^i_y {\int} \frac{dp^-}{(p^-)^3}
  \frac{1}{4\pi} \big[\theta (x^- - y^-)\theta (p^-)
  - \theta (y^- -x^-)\theta (-p^-) \big] \\
  \times \int d^2 z_{\perp} \frac{d^2 p_{\perp}}{(2\pi)^2}\frac{d^2
    q_{\perp}}{(2\pi)^2} e^{i p_{\perp} (x_{\perp} -
    z_{\perp})+iq_{\perp} (z_{\perp} - y_{\perp})}
  e^{-i\frac{{p_{\perp}}^2}{2 p^-} x^-
    +i \frac{{q_{\perp}}^2}{2 p^-} y^-} \\
  \times \{\theta(x^-)\theta(y^-)+\theta(-x^-)\theta(-y^-) \\
  +\theta(-x^-)\theta(y^-) \tilde{U}
  (z_{\perp})+\theta(x^-)\theta(-y^-) \tilde{U}^{\dagger}
  (z_{\perp})\}\, .
\label{correlator}  
\end{multline} 
This expression displays a separation into $x^-$ ordered contributions
that seamlessly matches up with what we have already seen for the
vertices in Eq.~(\ref{expansion}).  Diagrammatically, the $x^-$
ordered exchange contributions are given by
\begin{align} 
  \label{eq:chidiagstimeordered} 
  \begin{minipage}[m]{2.5cm} \epsfysize=2.5cm  
    \epsfbox{chiqqb.ps}
  \end{minipage} 
  & =
  \begin{minipage}[m]{2.5cm} \epsfysize=2.5cm  
    \epsfbox{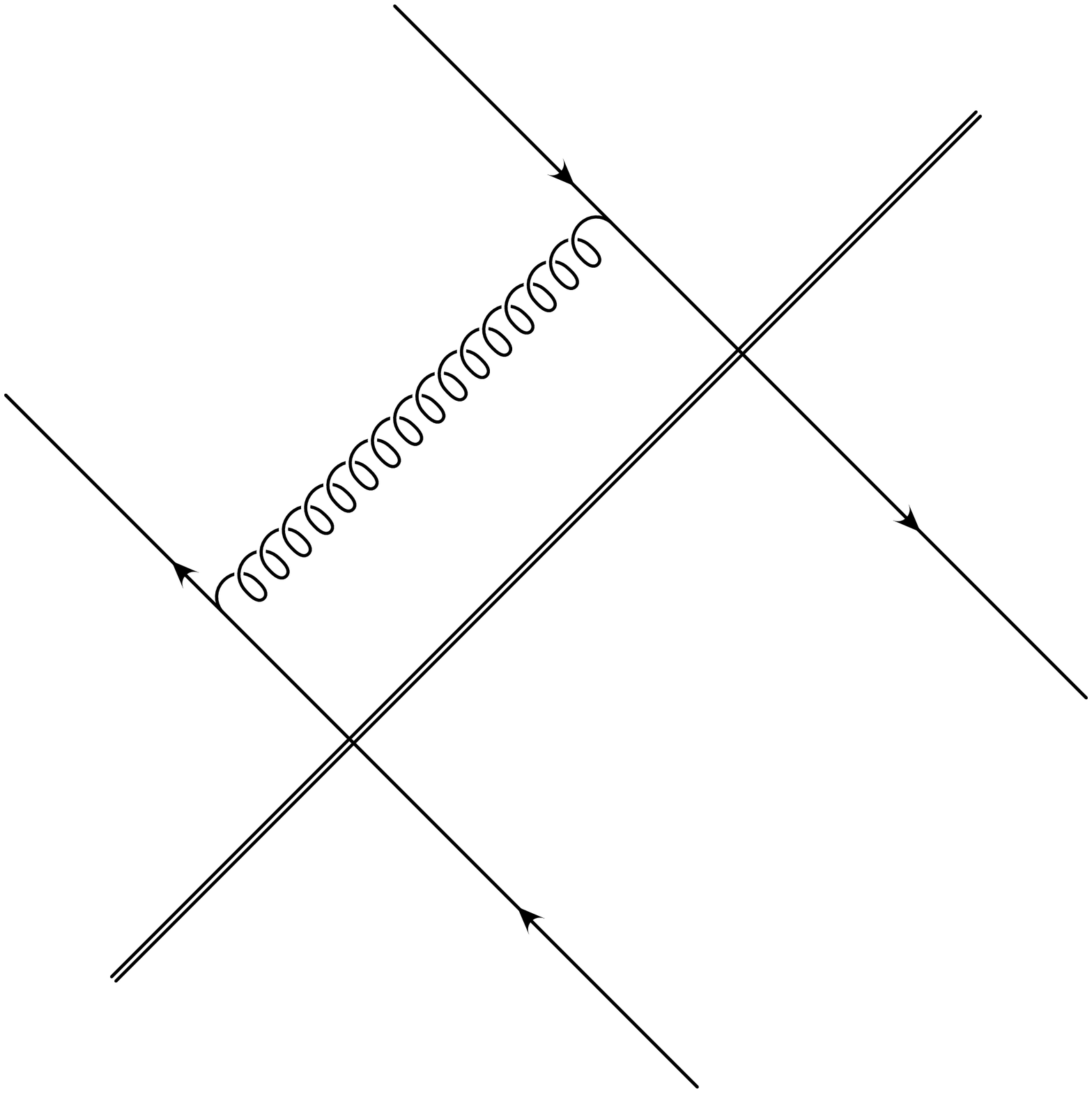}
  \end{minipage}  
  + \begin{minipage}[m]{2.5cm} \epsfysize=2.5cm \epsfbox{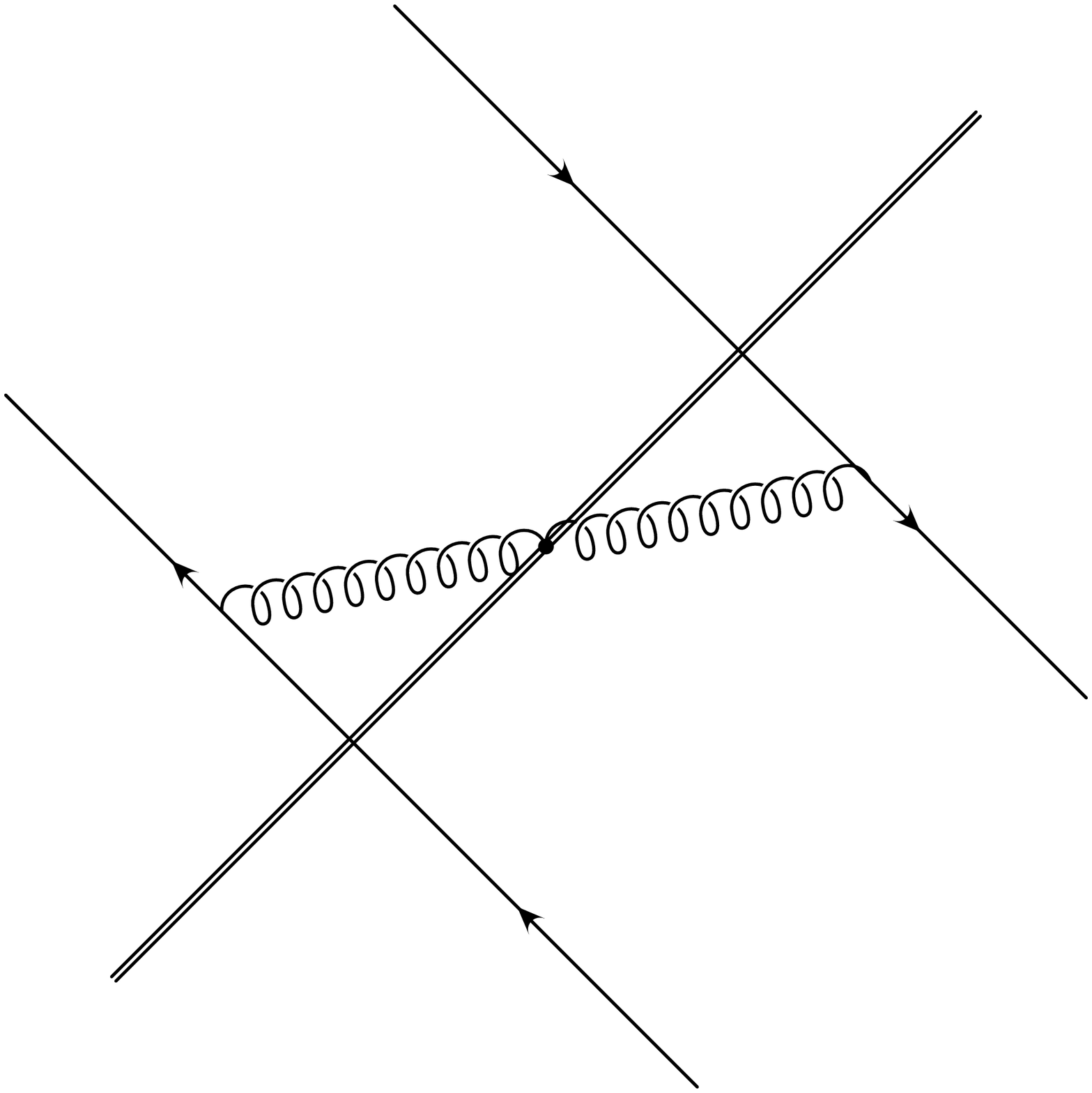}
  \end{minipage}  
  + \begin{minipage}[m]{2.5cm} \epsfysize=2.5cm \epsfbox{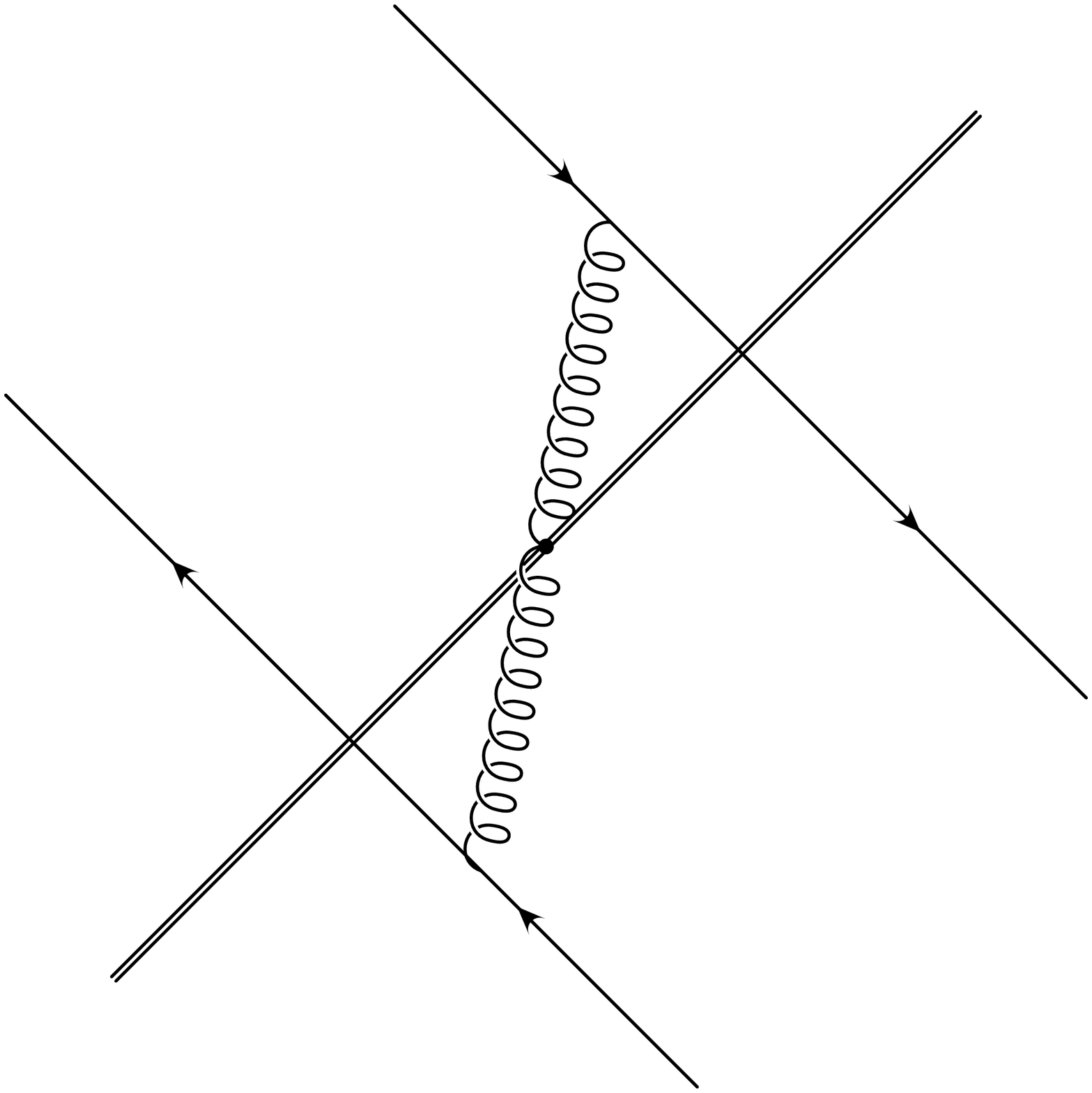}
  \end{minipage}  
  +\begin{minipage}[m]{2.5cm} \epsfysize=2.5cm
    \epsfbox{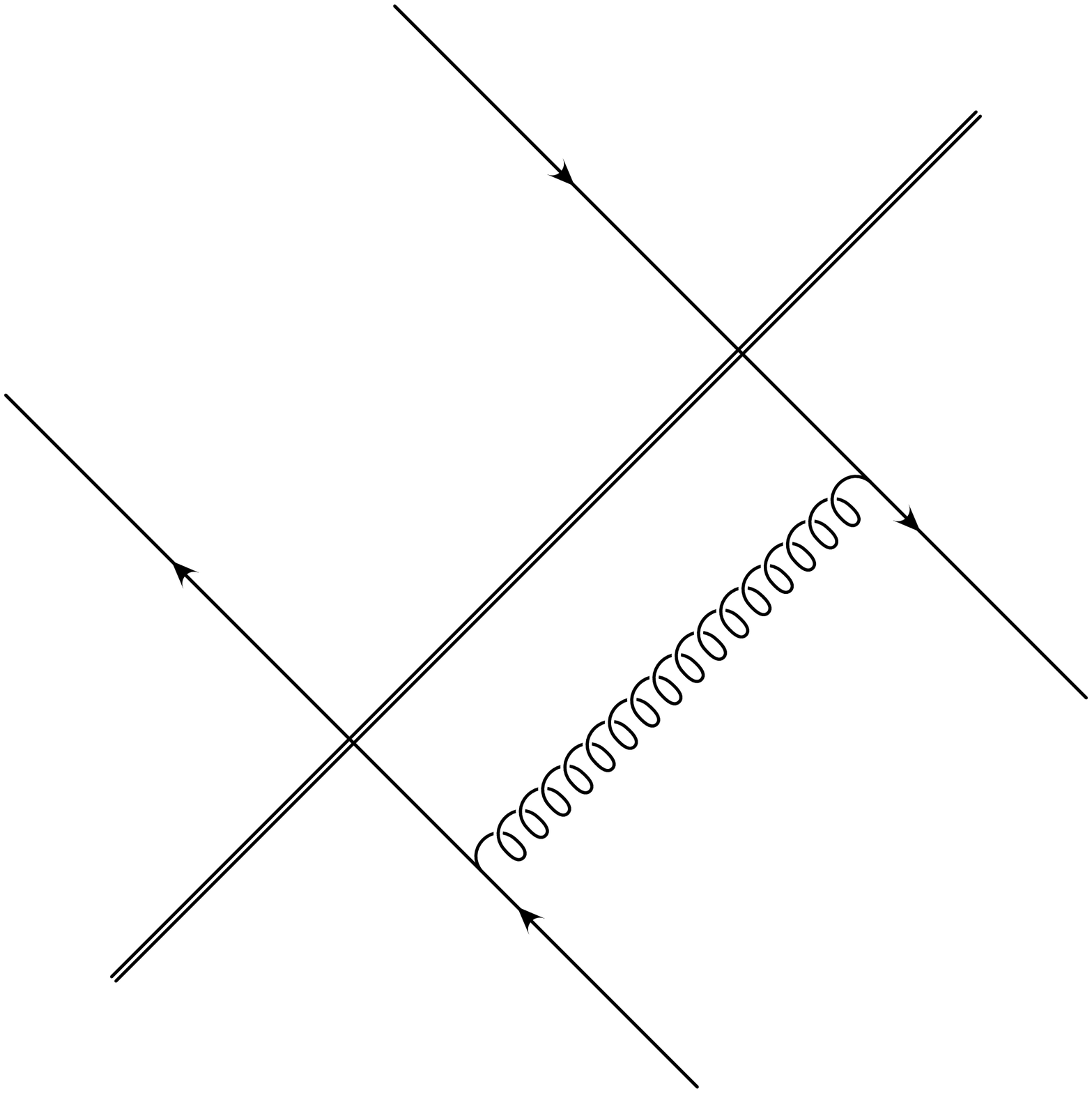}
  \end{minipage}\, ,      
\end{align} 
while the (connected parts of the) non-exchange ones are represented
by
\begin{subequations} 
  \label{eq:sigmadiags} 
  \begin{align} 
    \begin{minipage}[m]{2cm} \epsfysize=2cm  
      \epsfbox{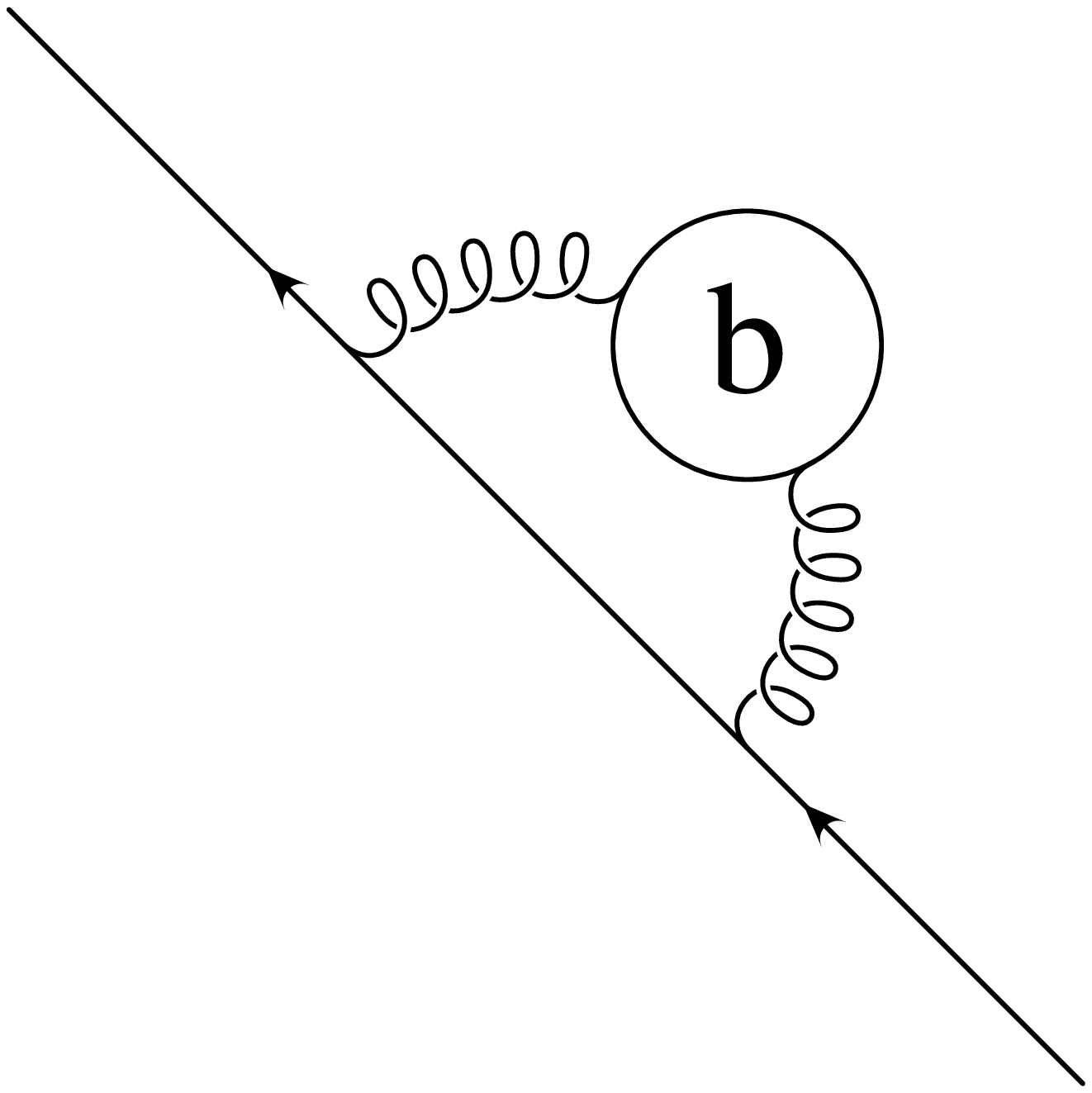}
    \end{minipage}   
    & =
    \begin{minipage}[m]{2.1cm} \epsfysize=2cm  
      \epsfbox{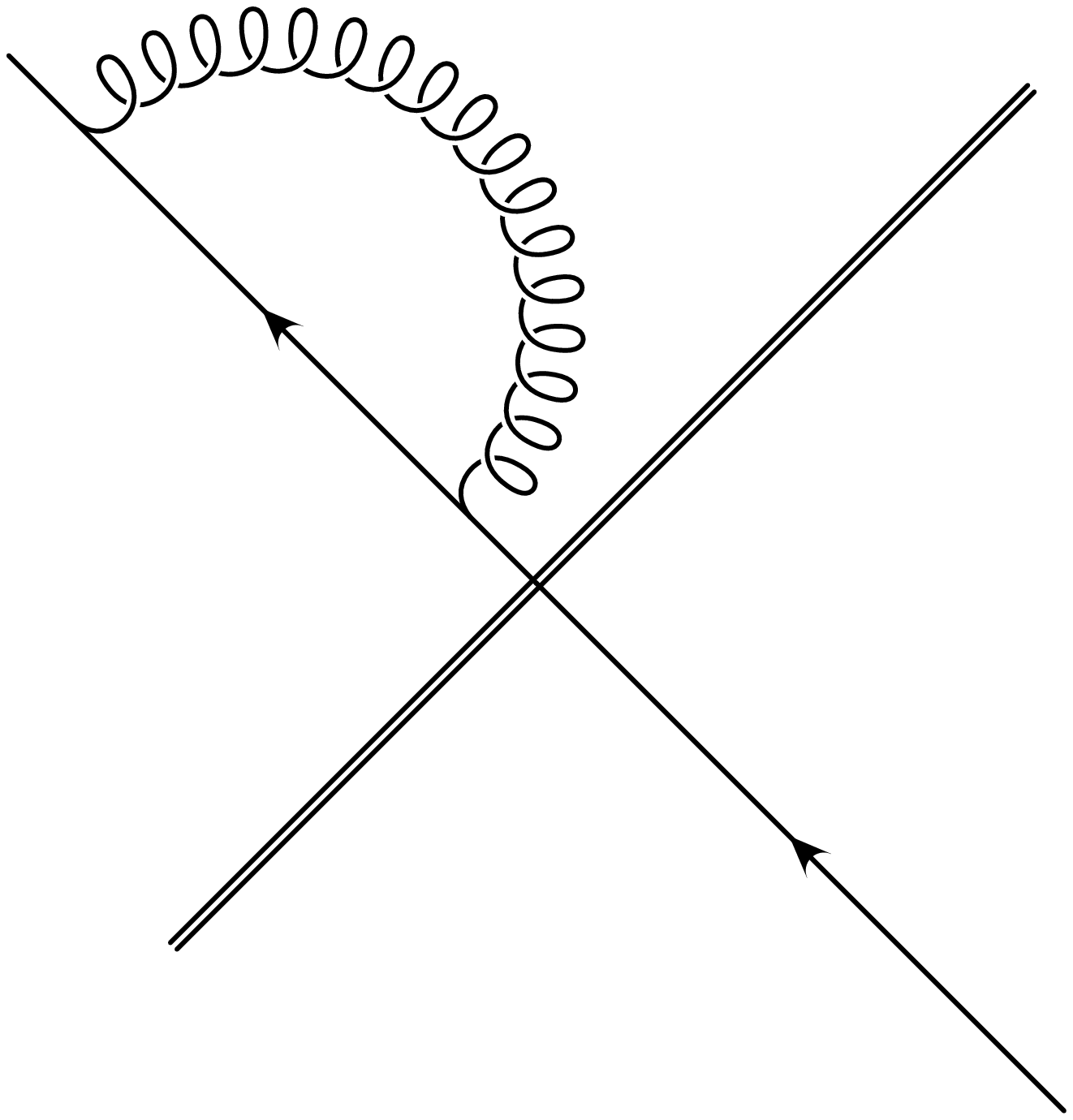}
    \end{minipage}   
    + \begin{minipage}[m]{2.1cm} \epsfysize=2cm
      \epsfbox{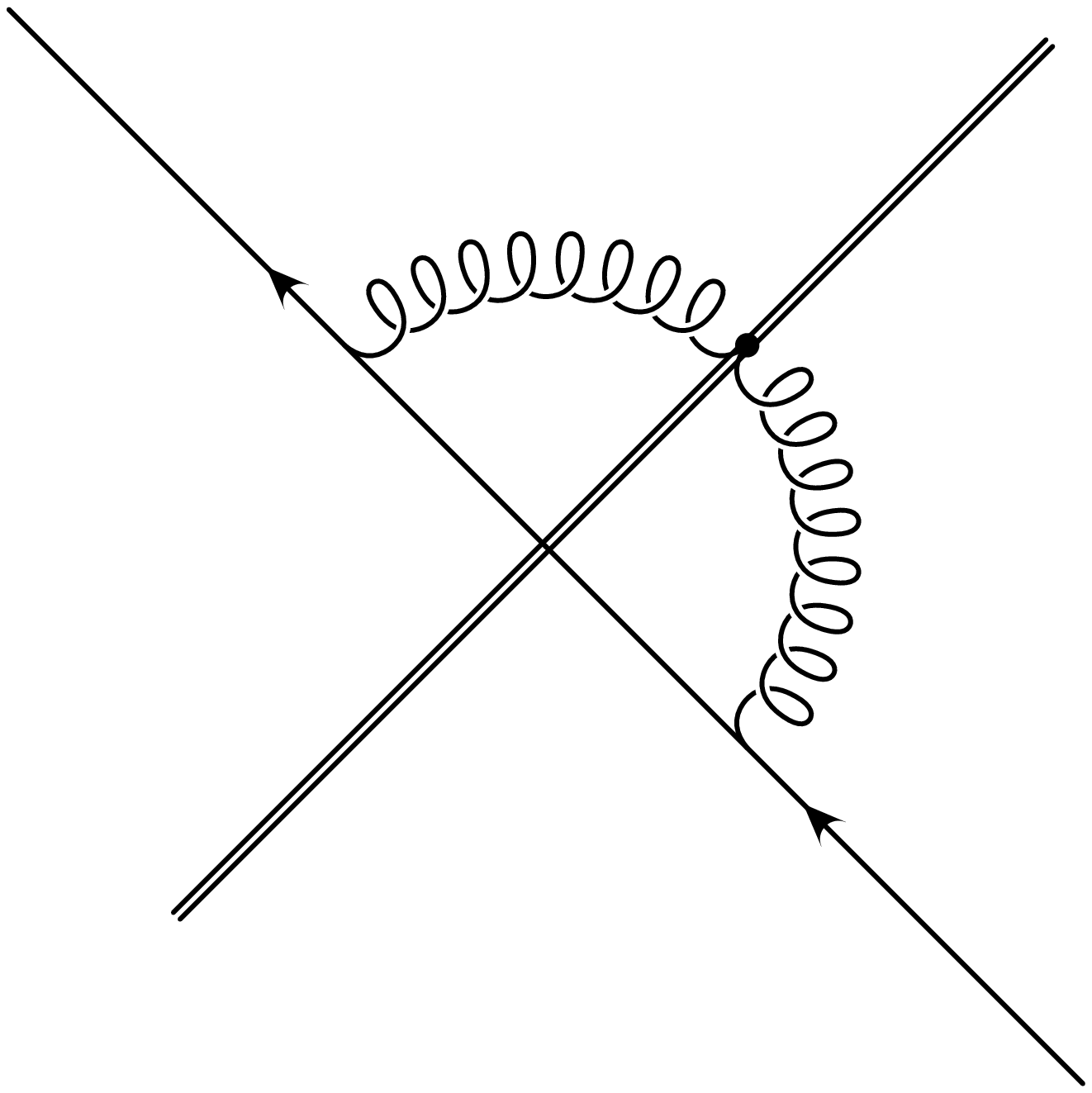}
    \end{minipage}   
    + \begin{minipage}[m]{2,1cm} \epsfysize=2cm
      \epsfbox{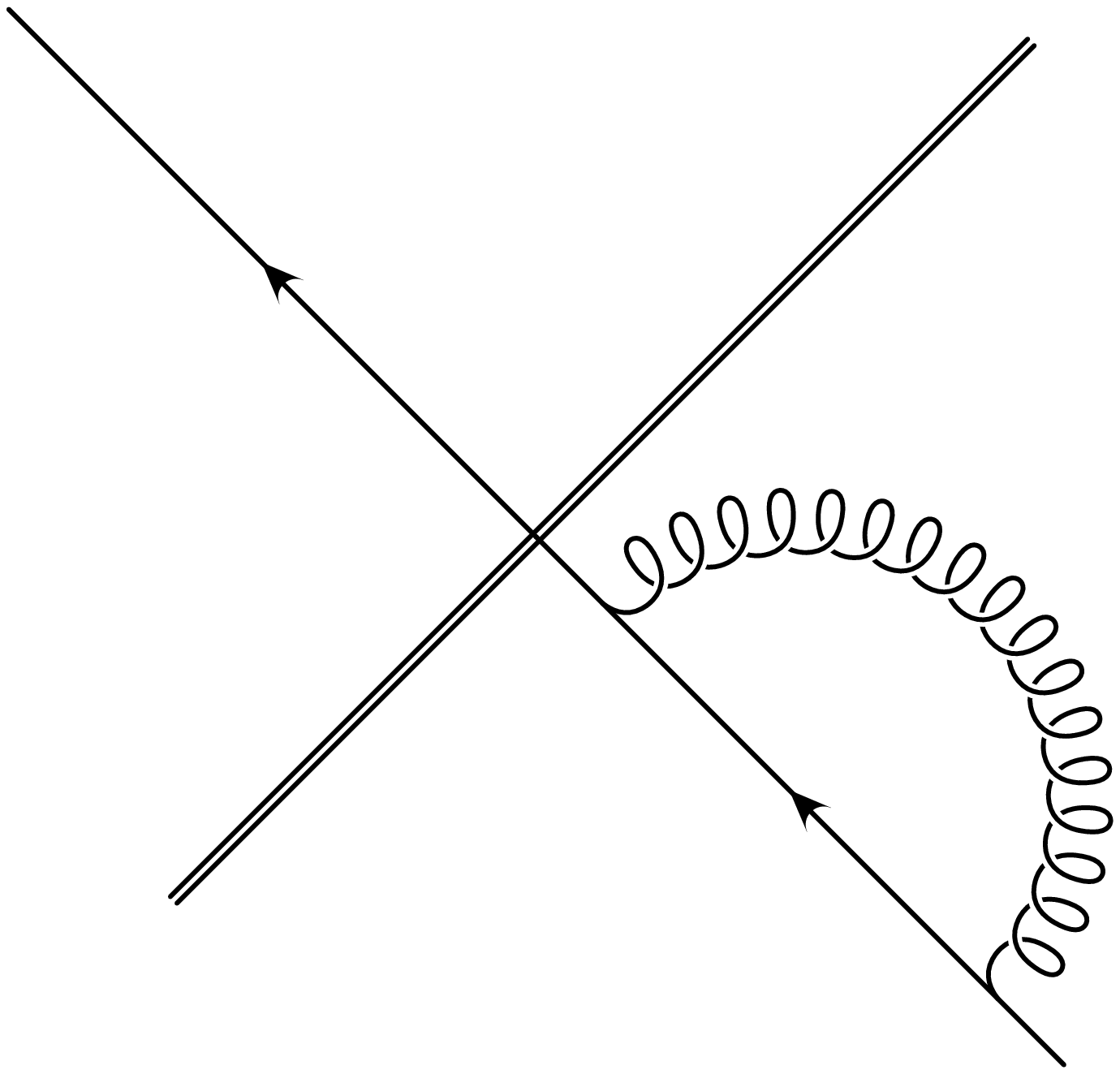}
    \end{minipage}\, ,    
    \\
    \begin{minipage}[m]{2cm} \epsfysize=2cm  
      \epsfbox{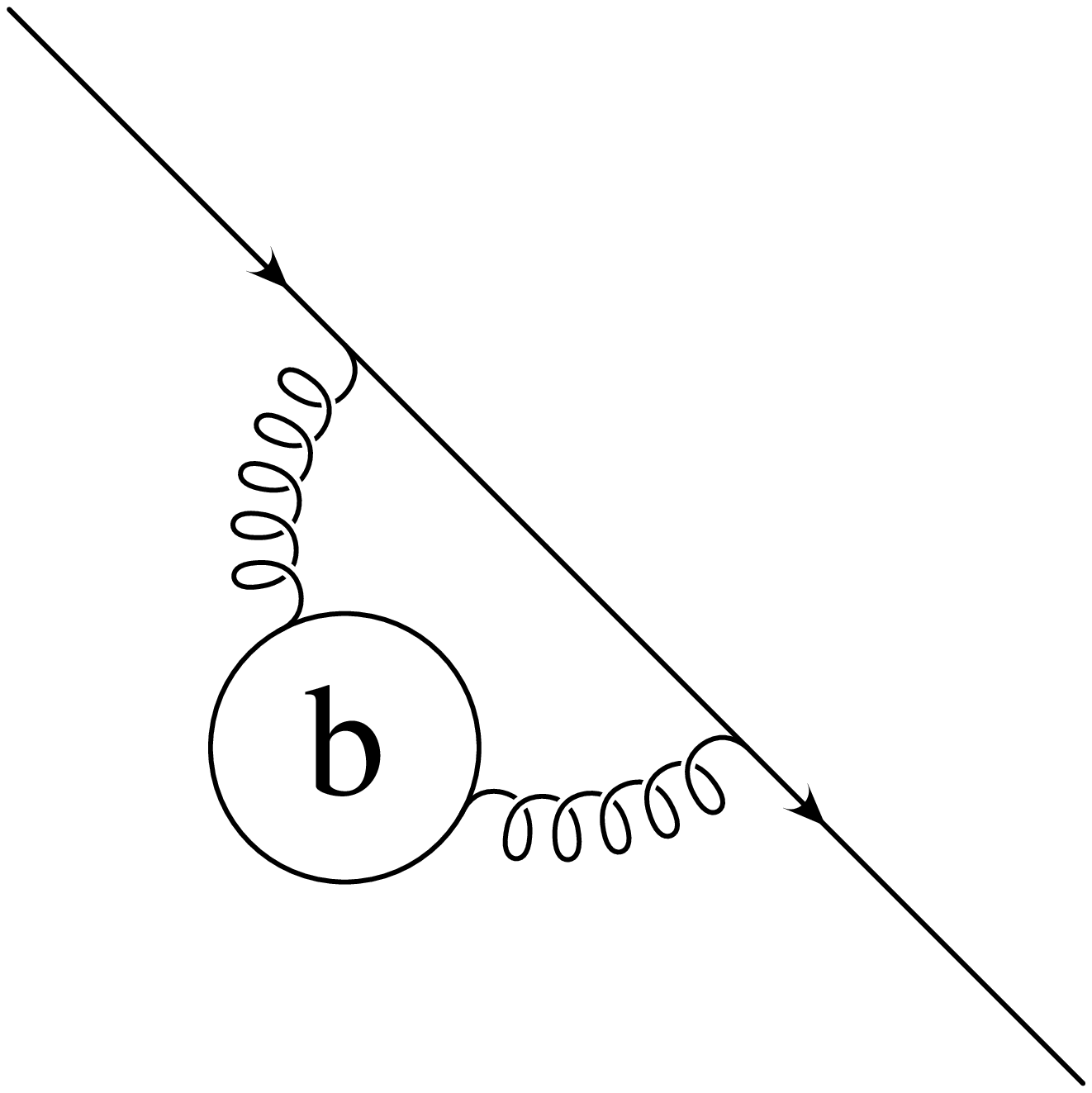}
    \end{minipage} 
    & =
    \begin{minipage}[m]{2.1cm} \epsfysize=2cm  
      \epsfbox{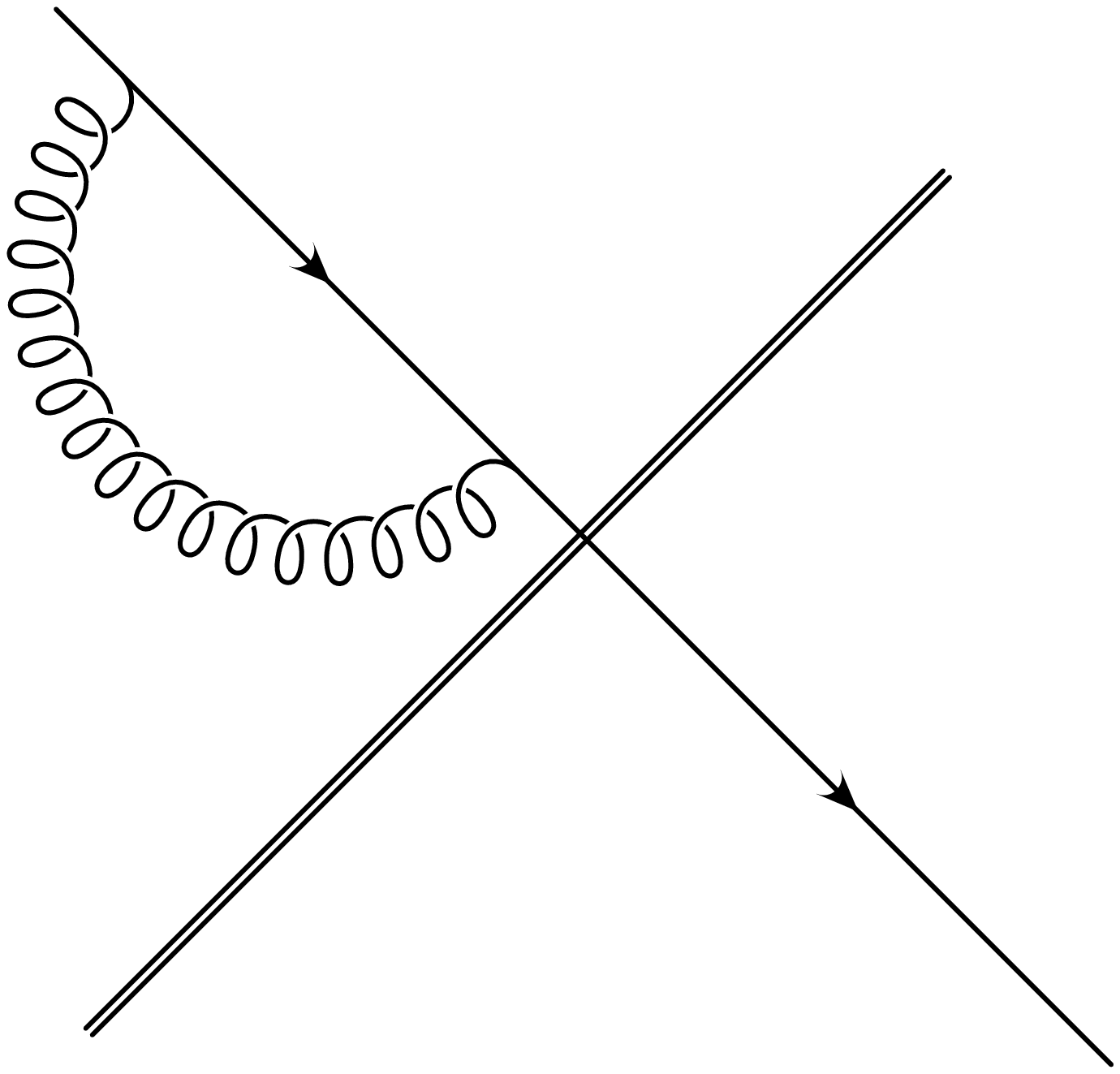}
    \end{minipage} 
    +
    \begin{minipage}[m]{2.1cm} \epsfysize=2cm  
      \epsfbox{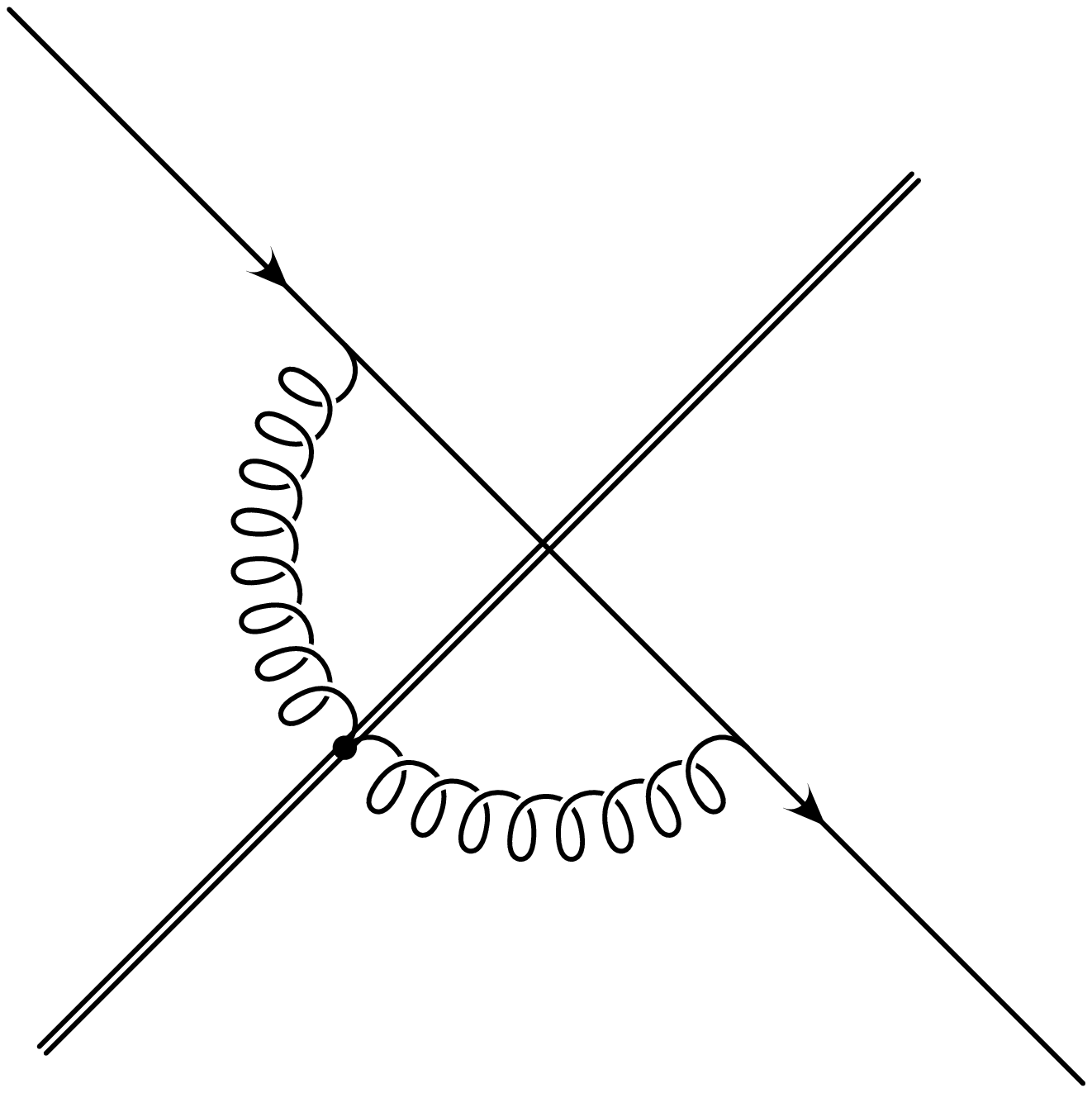}
    \end{minipage} 
    +
    \begin{minipage}[m]{2.1cm} \epsfysize=2cm  
      \epsfbox{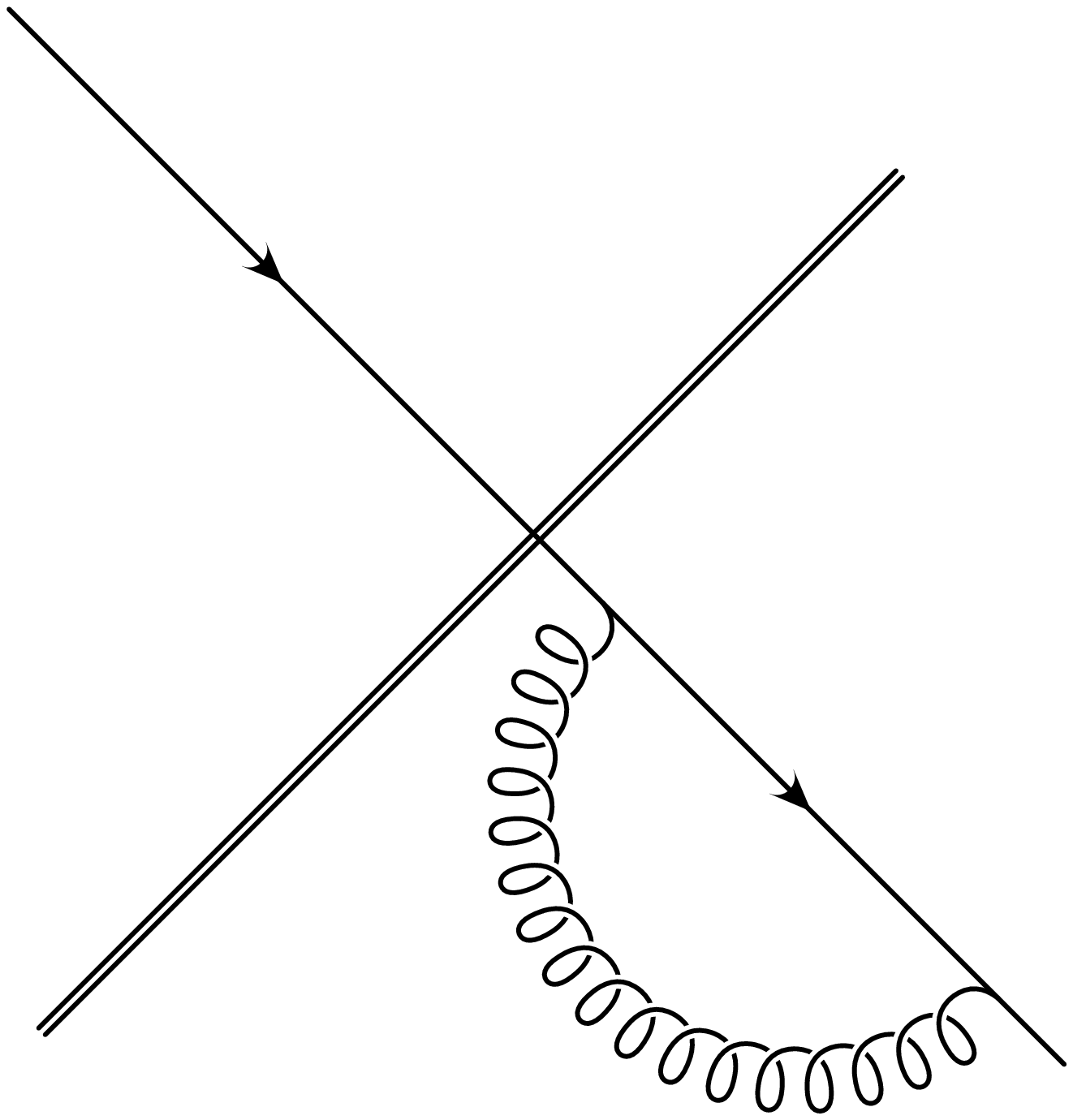}
    \end{minipage}\, . 
  \end{align} 
\end{subequations}

Whenever the $x^-=0$ plane cuts the fluctuation propagator, the
diagram contains a factor $U_{ab}^{(\dagger)}$, otherwise the
fluctuation propagator is free.

Algebraically, the corrections to the scattering cross section involve
integrals of this propagator with respect to $x^-$ and $y^-$ from zero
to either $+$ or $-$ infinity. This is straightforward to do.  For
example, for the second term in Eq.~(\ref{expansion}) we need
\begin{equation} 
  \begin{split} 
    \int\limits_{-\infty}^{0}\!\! dw^- \int\limits_{0}^{+\infty}\!\!
    dz^-
    \big\langle a_a^+ &(x_\perp , w^-) a_b^+ (y_\perp , z^-) \big\rangle = \\
    & =  {\partial}_{x}^{i} {\partial}_{y}^{i}
    \int\limits_{0}^{+\infty}\! \frac{dp^-}{p^-} \frac{1}{\pi} \int
    d^2 z_{\perp} \int \frac{d^2 p_{\perp}}{(2\pi)^2}\frac{d^2
      q_{\perp}}{(2\pi)^2} \frac{e^{i p_{\perp} (x_{\perp} -
        z_{\perp})+ i q_{\perp} (z_{\perp} -
        y_{\perp})}}{{p_{\perp}}^2 {q_{\perp}}^2}
    {\tilde{U}}_{ab} (z_\perp) \\
    & =  \frac{1}{\pi} \int\limits^{+\infty}_{0}\!
    \frac{dp^-}{p^-} \langle x_\perp|
    \frac{\partial^i}{\partial_\perp^2} {\tilde{U}}_{ab}
    \frac{\partial^i}{\partial_\perp^2}|y_\perp\rangle\, .
  \end{split} 
\end{equation} 
Here, $\langle x_\perp|O|y_\perp\rangle $ means the matrix element of
the operator $O$ in the coordinate basis in the usual sense. We treat
$\tilde U$ as an operator in the coordinate space with matrix elements
$\langle x_\perp|\tilde U|y_\perp\rangle =\tilde
U(x_\perp)\delta(x_\perp-y_\perp)$ and the products in the last line
are understood in the operatorial sense.  Explicitly,
\begin{equation} 
\label{exchange1} 
  \int\limits_{-\infty}^{0}\!\! dw^- \int\limits_{0}^{+\infty}\!\! dz^-  
  \big\langle a_a^+ (x_\perp , w^-) a_b^+ (y_\perp, z^-) \big\rangle  
  =  \frac{1}{4 \pi^3} \int\limits^{+\infty}_{0}\!  \frac{dp^-}{p^-}  
  \int d^2 z_{\perp} 
  \frac{(x-z)_\perp \cdot (y-z)_\perp}{(x-z)_\perp^2 (y-z)_\perp^2}  
  \cdot {\tilde{U}}_{ab}(z_\perp). 
\end{equation} 
In the same way we obtain for the other contributions with interaction
with the background
\begin{multline}  
\label{exchange2} 
\int\limits_{0}^{+\infty}\!\!  dw^- \int\limits_{-\infty}^{0}\!\!
dz^- \big\langle a_a^+ (x_\perp , w^-) a_b^+
(y_\perp , z^-) \big\rangle = \\
= \frac{1}{\pi} {\int}^{+\infty}_{0} \frac{dp^-}{p^-} \langle
x_\perp| \frac{\partial^i}{\partial_\perp^2} {\tilde{U}^\dagger}_{ab}
\frac{\partial^i}{ \partial_\perp^2}|y_\perp\rangle 
  = \frac{1}{4
  \pi^3} {\int}^{+\infty}_{0} \frac{dp^-}{p^-} \int d^2 z_{\perp}
\frac{(x-z)_\perp \cdot (y-z)_\perp}{(x-z)_\perp^2 (y-z)_\perp^2}
\cdot {\tilde{U}^\dagger}_{ab} (z_\perp)
\end{multline} 
\begin{multline}  
  - \int\limits_{-\infty}^{0}\!\! dw^- \int\limits_{0}^{+\infty}\!\!
  dz^- \big\langle a_a^+ (x_\perp , w^-) a_b^+
  (x_\perp , z^-) \big\rangle = \\
  = -\frac{1}{\pi} {\int}^{+\infty}_{0} \frac{dp^-}{p^-} \langle
  x_\perp| \frac{\partial^i}{\partial_\perp^2} {\tilde{U}}_{ab}
  \frac{\partial^i}{ \partial_\perp^2}|x_\perp\rangle = -\frac{1}{4
    \pi^3} {\int}^{+\infty}_{0} \frac{dp^-}{p^-} \int d^2 z_{\perp}
  \frac{1}{(x-z)_\perp^2} \cdot {\tilde{U}}_{ab} (z_\perp)
\label{nonexchange1} 
\end{multline} 
\begin{multline}  
  - \int\limits_{0}^{+\infty}\!\!  dw^- \int\limits_{-\infty}^{0}\!\!
  dz^- \big\langle a_a^+ (y_\perp , w^-) a_b^+
  (y_\perp , z^-) \big\rangle = \\
  = -\frac{1}{\pi} {\int}^{+\infty}_{0} \frac{dp^-}{p^-} \langle
  y_\perp| \frac{\partial^i}{\partial_\perp^2}
  {\tilde{U}^\dagger}_{ab} \frac{\partial^i}{
    \partial_\perp^2}|y_\perp\rangle =-\frac{1}{4 \pi^3}
  {\int}^{+\infty}_{0} \frac{dp^-}{p^-} \int d^2 z_{\perp}
  \frac{1}{(y-z)_\perp^2} \cdot {\tilde{U}^\dagger}_{ab} (z_\perp)
\label{nonexchange2} 
\end{multline}

To simplify the color structure of these expressions we use the
identity
\begin{align} 
  & {\tilde{U}}_{ab}(z_\perp) (t_a U(x_\perp ))^{\alpha\beta} (t_b
  U^\dagger (y_\perp ))^{\gamma\delta} = 2 \tr\big[t_a U(z_\perp ) t_b
  U^\dagger (z_\perp ) \big] (t_a U(x_\perp ))^{\alpha\beta} (t_b
  U^\dagger (y_\perp ))^{\gamma\delta}
  \nonumber\\
  & \qquad \quad= \frac{1}{2 N_c} \Big[ N_c {\big( U(z_\perp )\cdot
    U^\dagger (y_\perp)\big)}^{\alpha\delta} {\big( U^\dagger (z_\perp
    ) \cdot U(x_\perp )\big)}^{\gamma\beta} - U(x_\perp
  )^{\alpha\beta} U^\dagger (y_\perp)^{\gamma\delta} \Big]\, .
\end{align} 
 
Note that the integral over the frequency $p^-$ logarithmically
diverges.  In fact, we have to integrate only over a finite interval
of frequencies. The gluon field modes of very low frequency have been
already included in the background field $b^+$ and, therefore, the
fluctuation fields at these low frequencies should not be considered.
The lower cutoff on the frequency of the modes that are being
integrated is inversely proportional to the initial value of $x_0$ at
which we start the evolution. The upper limit on the high frequency
side is furnished by the maximal rapidity of the quark (or antiquark)
in the virtual photon which is of the order of $1/x$.  The ratio of
these two cutoffs is of order $x_0/x$.  Thus, in the leading
logarithmic approximation we identify
\begin{equation} 
  {\int}\frac{dp^-}{p^-} =\ln \frac{x_0}{ x}. 
\end{equation} 
 
The calculation of the remaining contributions (the ones with no
interaction with the background) proceeds along similar lines and is
given in the appendix.

Collecting all contributions together we obtain
\begin{align} 
\label{evolution} 
& \tr{\big\langle V(x_\perp ) V^\dagger (y_\perp ) \big\rangle_A } -
\tr{\big\langle U(x_\perp ) U^\dagger (y_\perp ) \big\rangle_b } =
\frac{g^2}{8 \pi^3} \ln (\frac{x_0}{x}) \int d^2 z_\perp  
\nonumber \\
& \hspace{.9cm} \times \bigg\langle\ \Big[ 2\tr{\big( U(x_\perp )
  U^\dagger (z_\perp)\big)} \tr{\big( U^\dagger (y_\perp ) U(z_\perp
  )\big)} - 2 N_c\tr{\big( U(x_\perp ) U^\dagger (y_\perp ) \big)}
\Big]
\frac{(x-z)_\perp{\cdot}(y-z)_\perp}{(x-z)_\perp^2 (y-z)_\perp^2} 
\nonumber \\
& \hspace{1.2cm} - \Big[ \tr\big( U(x_\perp ) U^\dagger (z_\perp)\big)
\tr{\big( U(z_\perp ) U^\dagger (y_\perp)\big)}
- N_c \tr(U(x_\perp )U^\dagger (y_\perp))\Big] \frac{1}{(x - z)^2_\perp} 
\nonumber \\
& \hspace{1.2cm} - \Big[ \tr \big( U^\dagger (y_\perp )
U(z_\perp)\big) \tr{\big( U^\dagger (z_\perp ) U (x_\perp)\big)} - N_c
\tr(U^\dagger (y_\perp )U (x_\perp)) \Big] \frac{1}{(y - z)^2_\perp} \ 
\ \bigg\rangle_b \, .
\end{align} 
The eikonal factors $V$ themselves should be considered as functions
of $x$, so that ${U=V(x_0)}$.  Differentiating this equation with
respect to $\ln 1/x$ we recover the evolution step for ${\tr \langle
  V(x_\perp) V^\dagger(y_\perp)\rangle_A}$.  This is precisely what
was found in Ref.~\cite{Balitskii:1996ub}.\footnote{As it stands,
  Eq.~(\ref{evolution}) does not provide a closed equation --- it has
  to be supplemented by evolution equations for arbitrary products
  ${\langle V^{(\dagger)}_1\otimes\ldots\otimes
    V^{(\dagger)}_n\rangle_A}$. The evolution of these higher
  correlators are derived following the same procedure as described
  above and leads to the full set of operator equations derived in
  Ref.~\cite{Balitskii:1996ub}. We will give a compact representation of the
  whole set of the evolution equations in Eq.~(\ref{final+}),
  Sec.~\ref{sect:more}.}
   
At large $N_c$ the products of traces in Eq.~(\ref{evolution})
factorize:
\begin{equation} 
  \langle\tr\big( U(x_\perp )\cdot U^\dagger (z_\perp)\big)  
  \tr\big( U^\dagger (y_\perp ) \cdot U(z_\perp )\big)\rangle 
  \xrightarrow{N_c\to\infty}  
  \langle\tr\big( U(x_\perp )\cdot U^\dagger (z_\perp)\big)\rangle  
  \langle\tr\big( U^\dagger (y_\perp ) \cdot U(z_\perp )\big\rangle. 
\end{equation} 
Equation (\ref{evolution}) then becomes a closed equation for the
evolution of ${N(x_\perp,y_\perp) =
  \big\langle\tr(U(x)U^\dagger(y)-1)\big\rangle}$.  It is identical to
the nonlinear evolution equation of Ref.~\cite{Kovchegov:1999yj}.
 
\subsection{The physical interpretation.}
Now, let us discuss the physical picture of this evolution. As always 
with DIS, the physical picture depends on the frame in which one 
chooses to view the process.  We have specified the frame to some 
extent by declaring that the photon fluctuates into a $q\bar q$ pair 
long before the target. However, we are still free to put the 
subsequent evolution in $x$ either into the evolution of the photon 
wave function or into the evolution of the gluon field distribution in 
the target.  We will refer to the former picture as the ``projectile 
evolution picture'' and to the latter as the ``target evolution 
picture''. The calculation is, fortunately, noncommittal on this point 
and we will consider both pictures in turn. 
 
In the projectile evolution picture, the higher energy of the 
scattering is achieved by boosting the $q\bar q$ pair.  In this 
picture, the quark and antiquark have very high energy and 
consequently their wave function develops extra gluon components.  The 
growth of the cross section with $1/x$ then is interpreted as due to 
the scattering of extra gluons in the projectiles wavefunction. This 
is precisely how the low $x$ evolution is viewed in the dipole model 
of Mueller \cite{mueller:1994rr,Mueller:1994jq}.  Our calculation of this section 
has a simple interpretation from this point of view.  The quark and 
the antiquark is the pair of pointlike color charges moving with 
velocity of light and located at $x^+=0$ and transverse coordinates 
$x_\perp$ and $y_\perp$.  These color charges carry with them a 
fluctuating gluon field. When the pair is boosted to higher rapidity 
the gluon fields ``freeze'' due to time the dilation and become static. In 
the approximation when the gluon fields are frozen, they are given by 
the Weisz\"acker-Williams static ($p^+=0$) fields created by the 
$q\bar q$ pair. In the leading order in $\alpha_s$ the 
Weisz\"acker-Williams (WW) fields are small and are emitted 
independently by the quark and the antiquark. The total WW field is 
\begin{equation} 
  A^i=g\frac{1}{ \partial^-}\frac{\partial^i}{ \partial_\perp^2}  
[j^-_q+j^-_{\bar q}]\, , 
\end{equation} 
where $j^-_q$ ($j^-_{\bar q}$) is the color current due to the quark 
(antiquark) which in our frame has only a ``$-$'' component. For 
pointlike quark and antiquark the charge densities are delta functions 
in the transverse coordinates and in $x^+$.  The WW field is therefore 
\begin{equation} 
  A^i(p^-,z_\perp)=g\frac{1}{ p^-}[\tau\frac{x^i-z^i}{ (x-z)^2}+  
\tau'\frac{y^i-z^i}{ (y-z)^2}]\, . 
\end{equation} 
Here, $\tau$ and $\tau'$ are fundamental color matrices corresponding 
to the orientation of the quark and antiquark wavefunctions in the 
color space. Their exact form does not matter for our purposes.  The 
WW field, if written in the particle basis, can be thought of as 
representing equivalent gluons.  The number of gluons at a given 
transverse position is given by the familiar expression 
\begin{align} 
  n_{WW}(z_\perp)&\propto \int_0^\infty dp^- p^- \, \tr 
  F^{-i}(p^-,z_\perp)F^{-i}(p^-,z_\perp) \nonumber \\ 
  &=\alpha_s \int_0^\infty \frac{dp^-}{ p^-} \, \tr 
  [\tau\frac{x^i-z^i}{ (x-z)^2}+ \tau'\frac{y^i-z^i}{ (y-z)^2}] 
  [\tau\frac{x^i-z^i}{ (x-z)^2}+ \tau'\frac{y^i-z^i}{ (y-z)^2}]\, . 
\end{align} 
If we do not take the trace over the color indices, this expression 
gives the probability to have one extra WW gluon in the wave function 
of the $q\bar q$ pair (at the transverse position $z_\perp$ with a 
particular color orientation).  These WW gluons scatter on the gluon 
field of the target eikonally just like the quark and the antiquark, 
apart from the fact that they carry adjoint charge, and so their 
eikonal amplitude is given by $\tilde U$ rather than $U$.  The 
terms in this expression are in one-to-one 
correspondence with the real
contributions in Eq.~(\ref{expansion}), that is the terms
in which the gluons interact with 
the target (background). 
The rest of the terms in Eq.(\ref{expansion}) - the virtual terms - as 
usual serve to restore the correct normalization of the wave function.
 
To summarize, in the projectile evolution picture our calculation 
describes emission of the WW gluons into the wave function of $q\bar 
q$ long before the scattering.  The transverse coordinates of these 
gluons are frozen due to the Lorentz time dilation.  Subsequently, 
both the $q$ and $\bar q$, and also the gluons, scatter eikonally and 
independently of each other on the target gluon field.  Clearly, this 
picture is identical to the dipole evolution picture of Mueller which 
was used in Ref.~\cite{Kovchegov:1999yj} to derive a nonlinear evolution equation. 
The only difference is that the dipole model uses the simplifications 
in the color algebra which arise in the large $N_c$ limit. 
 
The calculation presented above also has a simple interpretation in
the target evolution picture.  In this picture, it is the target
rather than the projectile that is boosted when going to lower $x$. As
already made explicit above by writing $\langle\ldots\rangle_b$, one
should think about the target as being represented by an ensemble of
the configurations of $b^+$.  The corresponding statistical weight
$Z[b]$ is determined, of course, by the structure of the target at the
relevant resolution scale. We will have more to say about it in
Sec.~\ref{sect:more}.  The boost of the target freezes the gluon field
fluctuations around the target background $b^+$ and, consequently,
some field modes which were not important at higher $x$ are now
capable of inducing scattering. Thus, the ensemble of the relevant
field configurations which characterizes the target changes.  In fact,
every $b^+$ now forks into a ``subensemble'' $b^{+'}=b^++a^+$. In the
weak coupling regime $a^+$ have Gaussian distribution with the width
determined by the inverse of their correlation function
Eq.~(\ref{correlator}).  One can work back from here and calculate the
modification of the distribution of the background fields. We will do
this in the following sections.
 
The fluctuations of $a^+$ are, therefore, considered in the target
evolution picture as modifying the ensemble of the target background
fields very much like in the approach of
Refs.~\cite{Jalilian-Marian:1997xn, Jalilian-Marian:1997jx,
  Jalilian-Marian:1997gr,Jalilian-Marian:1997dw, Kovner:1999bj}.
 
\subsection{Unitarization in different approximations.}
{}From what has been said so far, it is clear that although the
calculation presented in this section includes into the evolution some
nonlinear effects, it is not the end of the story. At very low $x$,
this approximation should break down.  There are clear reasons why
this should happen in both pictures.  In the projectile evolution
picture, it is not true indefinitely that the WW fields are emitted
independently from the partons in the projectile.  Due to the
evolution, more and more gluons are emitted into the wave function of
the projectile and so the density of partons grows. At some point, the
approximation of independent emissions as well as of independent
scattering of the partons on the target must break down. This is the
point at which, in the parlance of Ref.~\cite{Kovchegov:1999yj}, the
pomeron loop diagrams must come into play.  In the target evolution
picture, the problematic point is the eikonal approximation for the
scattering of the $q\bar q$ pair. At not very low $x$ the target
fields are not too strong. Since the $q\bar q$ pair is very energetic,
the eikonal approximation is perfectly valid. However, with the
evolution the strength of the target fields grows. The energy of the
$q\bar q$ pair, on the other hand, stays fixed. When the fields are
strong enough the quark and antiquark will start losing a finite
fraction of their energy and, therefore, the no recoil eikonal
approximation cannot stay valid indefinitely.
 
The effect of the nonlinear evolution Eq.~(\ref{evolution}) on the 
behavior of the total cross section was studied in 
Refs.~\cite{Kovchegov:1999ua,Levin:1999mw}. It was concluded that the 
nonlinearities slow down the BFKL type rise of the cross section and 
lead to its unitarization so that the cross section approaches the 
black disk limit. We want to conclude this section with a comment on 
the nature of the unitarization in this approximation. Essentially, 
the unitarization is brought about by purely kinematical effects. This 
is especially clear in the projectile evolution picture. At the 
initial value of $x=x_0$ one starts with the $q\bar q$ pair as the 
only relevant component of the photon wave function, which has a 
certain probability $P_{q\bar q}$ to scatter on the target.  So, 
initially, the total scattering probability $P_{x_0}$ is 
\begin{equation} 
  P_{x_0}=P_{q\bar q}\, . 
\end{equation} 
At lower $x=x_0-\delta x$ the wave function also contains a component 
with an extra gluon. Let the probability to have an extra gluon in the 
wave function be $\Delta$ and the probability for this gluon to 
scatter on the target $P_g$.  In the linear approximation (the BFKL 
limit) the total probability of scattering is additive 
\begin{equation} 
  P_{x}=(1-\Delta)P_{q\bar q}+\Delta(P_{q\bar q}+P_g)=  
P_{q\bar q}+\Delta P_g \, . 
\end{equation} 
However, this is, in fact, overcounting, since there are events where 
both the gluon and the $q\bar q$ pair undergo scattering and those 
events are counted twice in the linear approximation. One should, 
therefore, subtract the probability of these double scattering events 
from the total probability. This deficiency is corrected by writing 
\begin{equation} 
  P_x=(1-\Delta)P_{q\bar q}+\Delta(P_{q\bar q}+P_g-P_{q\bar q}P_g)=  
P_{q\bar q}+\Delta(P_g-P_{q\bar q}P_g) =P_{x+\delta 
     x}+\Delta(1-P_{x+\delta x})P_g\, . 
\end{equation} 
At arbitrary low $x$, the same argument leads to a similar expression 
where $P_{x+\delta x}$ denotes the total scattering probability of the 
projectile (which itself contains the $q\bar q$ pair and some number 
of gluons) at a slightly higher value of $x$.  This is precisely the 
nonlinear term in the evolution equation Eq.~(\ref{evolution}) with 
the only difference that the extra gluon in the wave function can have 
arbitrary transverse coordinate and one should, of course, integrate 
over this extra degree of freedom.  It is clear that this negative 
nonlinear correction leads to the unitarization of the cross section 
since as $P_x$ tends to unity the emission of the extra gluon does not 
increase the total scattering probability.  This effect is somewhat 
similar to the Glauber mechanism, not in the sense that each parton 
undergoes multiple scattering, but that the unitarization is of a 
purely geometrical nature. A similar discussion, in the framework of 
the dipole model, is given in Refs.~\cite{Mueller:1995gb,Kovchegov:1997dm}. 
 
In the next section we will show how to relate the approach just 
discussed with that of 
Refs.~\cite{Jalilian-Marian:1997xn,Jalilian-Marian:1997jx,Jalilian-Marian:1997gr,Jalilian-Marian:1997dw,Kovner:1999bj}.

\section{The JKLW equation and the small induced field limit.} 
\label{sect:jklw}

We start this section by recalling the framework and results of 
Refs.~\cite{Jalilian-Marian:1997xn,Jalilian-Marian:1997jx,Jalilian-Marian:1997gr,Jalilian-Marian:1997dw,Kovner:1999bj,Jalilian-Marian:1998cb}.  

\subsection{The JKLW equation.}
In this approach, following Refs.~\cite{mclerran:1994ni,mclerran:1994ka}, the averages of
gluonic observables in a hadron are calculated via the following path
integral\footnote{An alternative form of the effective action was
  suggested in~\cite{Jalilian-Marian:2000ad} where it was also shown
  that it leads to the same evolution equation.}
\begin{multline} 
  \big\langle O(A)\big\rangle =\int D\alpha^i DA^\mu O(A)  
  Z[\alpha
  ] \times \\ 
  \times \exp\bigg\{ -i\int d^4 x \frac{1}{ 4}{\tr} 
  F^{\mu\nu}F_{\mu\nu} -\frac{1}{N_c} \int d^2 x_\perp dx^- \delta 
  (x^-) \partial^i\alpha_a^i(x_\perp) {\tr}T_a {\cal 
    W}_{-\infty,\infty} [A^-](x^-,x_\perp)\bigg\}\, , 
\label{actionf} 
\end{multline} 
where the gluon field strength tensor is given by 
\begin{equation} 
  F^{\mu\nu}_{a} =  
{\partial}^{\mu} A^{\nu}_{a} - {\partial}^{\nu} A^{\mu}_{a}  
- g f_{abc}A^{\mu}_{b}A^{\nu}_{c} 
\end{equation} 
and ${\cal W}$ is the Wilson line in the adjoint representation along 
the $x^+$ axis 
\begin{equation} 
  {\cal W}_{-\infty ,+\infty}[A^{-}](x^- , x_{\perp}) =  
  {\cal P}\exp\Bigg[+i g \int dx^+ A^-_a (x^+,x^- , x_{\perp}) T_a  
  \Bigg]\, . 
\end{equation} 
 
The hadron is represented by an ensemble of chromoelectric fields, 
localized in the plane $x^-=0$, of the form 
\begin{equation} 
  f^{+i}=\frac{1}{ g}\delta(x^-)\alpha^{i}(x_\perp)\, , 
\label{chrom} 
\end{equation} 
where the two dimensional vector potential $\alpha^i(x_\perp)$ is 
``pure gauge'' 
\begin{align} 
  {\partial}^{i}{\alpha}_{a}^{j} &- {\partial}^{j}{\alpha}_{a}^{i} - 
  f_{abc}{\alpha}_{b}^{i}{\alpha}_{c}^{j}=0\, . 
\label{sol} 
\end{align} 
In Eq.~(\ref{actionf}), $Z[\alpha]
$ is the statistical weight of a configuration $\alpha_i(x_\perp)$ in 
the hadronic ensemble. 
 
The evolution in Refs.~\cite{Jalilian-Marian:1997jx,Jalilian-Marian:1997gr,Jalilian-Marian:1997dw,Kovner:1999bj} is 
derived in the target evolution picture where decreasing $x$ 
corresponds to boosting the hadronic target.  This leads to freezing 
of part of the gluonic degrees of freedom.  Integrating out these slow 
modes of the vector potential generates the renormalization group 
equation, which has the form of the evolution equation for the 
statistical weight $Z$ \cite{Jalilian-Marian:1997xn,Jalilian-Marian:1997jx,Jalilian-Marian:1997gr,Jalilian-Marian:1997dw,Kovner:1999bj} 
\begin{equation} 
  \frac{d}{ d\ln \frac{1}{x}}Z= \alpha_s \Big\{\frac{1}{ 2}  
  \frac{\delta^2}{\delta\alpha(u)\delta\alpha(v)} 
    \big[Z\chi(u, v) \big] - \frac{\delta}{\delta\alpha(u)} 
    \big[Z\sigma(u)\big]\Big\}\, . 
\label{final} 
\end{equation} 
In the compact notation used in Eq.~(\ref{final}), both $u$ and $v$ 
stand for color and rotational index and transverse coordinates, with 
summation and integration over repeated occurrences implied.  This 
evolution equation for the statistical weight can be rewritten as the 
set of the evolution equations for the correlation functions of the 
chromoelectric field 
\begin{multline} 
\label{correlf} 
\frac{d}{ d\ln\frac{1}{x}} 
\big\langle \alpha_{a_1}^{i_1}(x_1)...\alpha_{a_n}^{i_n}(x_n)\big\rangle  = \\ 
= \alpha_s \Bigg[\sum_{0<l<n+1} <\alpha_{a_1}^{i_1}(x_1)... 
\alpha_{a_{l-1}}^{i_{l-1}}(x_{l-1}) 
\alpha_{a_{l+1}}^{i_{l+1}}(x_{l+1})... 
\alpha_{a_n}^{i_n}(x_n)\sigma_{a_l}^{i_l}(x_l)\big\rangle  \\ 
+ \sum_{0<m<k<n+1}<\alpha_{a_1}^{i_1}(x_1)... 
\alpha_{a_{m-1}}^{i_{m-1}} 
(x_{m-1})\alpha_{a_{m+1}}^{i_{m+1}}(x_{m+1})... \\ 
\times \alpha_{a_{k-1}}^{i_{k-1}}(x_{k-1}) 
\alpha_{a_{k+1}}^{i_{k+1}}(x_{k+1})... 
\alpha_{a_n}^{i_n}(x_n)\chi_{a_ma_k}^{i_mi_k}(x_m, x_k)\big\rangle 
\Bigg]\, . 
\end{multline} 
The quantities $\chi[\alpha]$ and $\sigma[\alpha]$ have the meaning of 
the mean fluctuation and the average value of the induced vector 
potential which arises from the field modes which become frozen due to 
extra boost of the hadronic target.  In the leading logarithmic 
approximation of Refs.~\cite{Jalilian-Marian:1997jx,Jalilian-Marian:1997gr,Jalilian-Marian:1997dw,Kovner:1999bj} the two 
quantities $\chi$ and $\sigma$ completely specify the low $x$ 
evolution.  We give here an explicit expression for the mean 
fluctuation $\chi$ which will be the focus of our interest throughout 
this section 
\begin{equation} 
  \chi_{ab}^{ij}(x_\perp,y_\perp)= 2\langle x_\perp| 
  \{\frac{D^i}{ D_\perp^2}[D_\perp^2-S^{-1}] \frac{D^j}{ 
     D_\perp^2}\}_{ab}|y_\perp\rangle\, . 
\label{chif} 
\end{equation} 
For convenience, we have defined 
\begin{align} 
  \alpha_{ab}^i &= f_{abc}\alpha^i_c \, , \nonumber \\ 
  D_{ab}^i &= \partial^i\delta_{ab}+\alpha^i_{ab}\, . 
\label{def} 
\end{align} 
The operator $S$ in Eq.~(\ref{chif}) is given by 
\begin{equation} 
  S = \frac{1}{ D_\perp^2}+2 [\frac{\partial^i}{\partial_\perp^2}- 
  \frac{D^i}{ D_\perp^2}][\frac{\partial^i}{\partial_\perp^2}- 
  \frac{D^i}{ D_\perp^2}]= 
  \frac{1}{ D_\perp^2}-2\frac{1}{ \partial_\perp^2} 
  \partial_\perp\alpha\frac{1}{ D_\perp^2} + 
  2\frac{1}{ D_\perp^2}D_\perp\alpha\frac{1}{\partial_\perp^2} 
\, . 
\label{s} 
\end{equation} 
 
\subsection{Where does it come from?}
Technically, these results are derived as follows. One considers the 
quantum corrections in the classical background field 
Eq.~(\ref{chrom}).  The calculation is performed in the lightcone 
gauge $A^+=0$ with the residual gauge fixing ${\partial^i 
  A^i(x^-\to-\infty)=0}$ which fixes the gauge completely.  In this 
gauge the chromoelectric field Eq.~(\ref{chrom}) corresponds to the 
background vector potential 
\begin{equation} 
  b^i=\theta(x^-)\alpha^{i}(x_\perp)\, . 
\label{back} 
\end{equation} 
Note that, as opposed to the previous section, here we are using a 
different lightcone gauge: $A^+=0$. As a consequence, the background 
vector potential has a different form. 
 
The complete set of on-shell small fluctuation solutions of the 
classical equations is 
\begin{multline} 
  a^{i}_{p^-,r} =  e^{ip^{-}x^{+}} \int d^{2}p_{\perp} \bigg[ \theta 
  (-x^-) \exp\left( i\frac{p^{2}_{\perp}}{ 2p^{-}}x^{-} - 
    ip_{\perp}x_{\perp}\right) v^{i}_{-, r}(p_{\perp}) \\ 
  + \theta (x^{-}) U (x_\perp ) \exp\left( i\frac{p^{2}_{\perp}}{ 
      2p^{-}}x^{-} - ip_{\perp}x_{\perp}\right) 
  \left[U^{\dagger}v^{i}_{+,r}\right](p_\perp ) + \theta (x^{-}) 
  \gamma^{i}_{+,r}\bigg]\, . 
\label{solut} 
\end{multline} 
Here, $r$ is the degeneracy label, which labels independent solutions 
with the frequency $p^-$.  In the free case it is conventionally 
chosen as the transverse momentum, $\{r\}=\{p_\perp\}$.  The matrix 
$U(x_\perp )$ is the $SU(N)$ matrix that parameterizes the two 
dimensional ``pure gauge'' vector potential $\alpha^i(x_\perp)$ 
\begin{equation*} 
  \alpha^i(x_\perp )= i U(x_\perp ) 
  \partial^i U^\dagger(x_\perp )\, . 
\end{equation*} 
The auxiliary functions $\gamma^i_+, v^i_\pm$ are all determined in 
terms of one vector function.  Choosing this independent function as 
$v^i_-$ we have 
\begin{align} 
  v^{i}_{+,r} &= \bigg[T^{ij} -L^{ij} \bigg]  
  \bigg[t^{jk} -l^{jk}\bigg] v^{k}_{-,r}\, , \\ 
  \gamma^{i}_{+,r} &= 2 D^{i} \bigg[\frac{D^j}{ D_\perp^2} - 
  \frac{\partial^j}{ \partial_\perp^{2}}\bigg] \bigg[t^{jk} 
  -l^{jk}\bigg]v^k_{-,r}\, , 
\label{gammav} 
\end{align} 
where we have defined the projection operators 
\begin{alignat}{2} 
  T^{ij} & \equiv \delta^{ij} - \frac{D^i D^j}{ D_\perp^2}\, , 
  & \qquad L^{ij} & \equiv \frac{D^i D^j}{ D_\perp^2}\, , \nonumber\\ 
  t^{ij} & \equiv \delta^{ij} - \frac{\partial^i \partial^j}{ 
    \partial_\perp^2}\, , & \quad l^{ij} & \equiv \frac{\partial^i 
    \partial^j}{ \partial_\perp^2}\, . 
\label{eq:tproj} 
\end{alignat} 
The $\gamma_+$ piece of the eigenfunction Eq.~(\ref{solut}) is 
responsible for the induced vector potential since this is the only 
contribution that does not vanish at $x^-\to\infty$, so that 
\begin{equation} 
  \chi_{ab}^{ij}(x_\perp,y_\perp)=4\pi\int d p^-  
  \big\langle  
  \gamma^{i}_{+,a}(x_\perp,p^-)\gamma^{j}_{+,b}(y_\perp,-p^-) 
  \big\rangle\, . 
\label{defchi} 
\end{equation} 
 
Note that the essential nonlinearity of the expression 
Eq.~(\ref{chif}) is due to the denominator in the operator $S^{-1}$ 
Eq.~(\ref{s}). The reason this arises is due to the nontrivial 
normalization of the small fluctuation eigenfunctions.  As discussed 
in detail in Refs.~\cite{Jalilian-Marian:1997jx,Jalilian-Marian:1997gr,Jalilian-Marian:1997dw,Kovner:1999bj} the proper 
normalization of the eigenfunctions requires $v^i_-$ to be chosen as a 
complete set of eigenfunctions of the two dimensional Hermitian 
operator $O^{-1}$ 
\begin{equation} 
  \qquad [(t-l)O^{-1}(t-l)]_{ab}^{ij}(x_\perp,y_\perp)=  
  \langle x_\perp|\delta_{ab}^{ij}  
  - 2 \bigg[[\partial^i\frac{1}{\partial_\perp^2} 
  -D^i\frac{1}{ D_\perp^2}] 
  S^{-1}[\frac{1}{\partial_\perp^2}\partial^j 
  -\frac{1}{ D_\perp^2}D^j] \bigg]_{ab}|y_\perp\rangle\, , 
\label{o} 
\end{equation} 
such that 
\begin{equation} 
  \int d^2 r_\perp v^{i}_{-,r,a}(x_\perp ) v^{\ast j}_{-,r,b}(y_\perp )  
  = \frac{1}{ 4\pi |p^-|}[O^{-1}]_{ab}^{ij}(x_\perp ,y_\perp )\, . 
\label{ort} 
\end{equation} 
 
This nontrivial normalization is the consequence of the presence of 
the $\gamma_+$ piece in the solution Eq.~(\ref{solut}).  Equation 
(\ref{defchi}), supplemented by Eq.~(\ref{gammav}) and the 
normalization Eq.~(\ref{ort}), leads to the final expression 
Eq.~(\ref{chif}). 
 
If the contribution of $\gamma_+$ could be neglected in the 
normalization condition, the normalization of the eigenfunctions would 
be trivial and we would have $O=1$ in Eq.~(\ref{ort}).  One can 
consider the limit in which $\gamma_+$, or equivalently $\chi$, is 
small.  In the leading order in the expansion in $\gamma_+$ we have a 
very simple expression for $\chi$ \footnote{If so desired this expression 
can be written in a simple form in terms of the unitary matrix $U$, since 
operatorially $D^i=U\partial^iU^\dagger$. In Fourier space this gives 
convolutions of $U(p)$ and powers of transverse momentum.} 
\begin{equation} 
  \tilde\chi_{ab}^{ij}(x_\perp,y_\perp)=  
  4\langle x_\perp|\Big[D^i \{\frac{1}{ \partial_\perp^2} 
  +\frac{1}{ D_\perp^2}- 
  \frac{1}{ \partial_\perp^2} \partial_\perp D_\perp 
  \frac{1}{ D_\perp^2}- \frac{1}{ D_\perp^2}D_\perp\partial_\perp 
  \frac{1}{ \partial_\perp^2}\}D^j\Big]_{ab}|y_\perp\rangle\, . 
\label{chif1} 
\end{equation} 
Note that this is a 
  different limit than the one in which the JKLW evolution reduces to 
  the BFKL equation \cite{Jalilian-Marian:1997jx,Jalilian-Marian:1997gr,Jalilian-Marian:1997dw,Kovner:1999bj}.  The BFKL 
  limit corresponds to the expansion in powers of the background field 
  $\alpha^i$.  Now, we are not assuming that $\alpha^i$ is small, but 
  rather that the correction induced by the evolution, $\gamma^i$ is 
  small. 
\subsection{BK to JKLW: transforming between the gauges.}
We will now see that Eq.~(\ref{chif1}) is reproduced precisely by 
translating the calculation of the previous section into the language 
of the JKLW evolution. 
 
In the previous section, following Ref.~\cite{Balitskii:1996ub}, we used the 
gauge $A^-=0$. This is a very convenient gauge from the point of view 
of the projectile evolution since the eikonal amplitudes in this gauge 
are given by simple Wilson line factors.  We will refer to this gauge 
as the ``projectile lightcone gauge'', or the ``projectile gauge'' for 
short.  The JKLW approach on the other hand uses the $A^+=0$ gauge, 
which is convenient for the target evolution picture since it 
simplifies the relation between the distribution functions and the 
correlators of the gluon fields. We will call this gauge the ``target 
lightcone gauge'' or, simply, the ``target gauge''.  Our immediate aim 
is, therefore, to calculate $\chi$ using the results of the 
calculation in the projectile gauge. 
 
To do this note that the relation between the fields in the target and 
projectile gauges is given by 
\begin{equation} 
  \frac{1}{g} B^\mu+A^\mu=V(\frac{1}{g}b^\mu+a^\mu)V^\dagger
  +\frac{i}{g}V\partial^\mu V^\dagger\, . 
\label{trans} 
\end{equation} 
To simplify the notation, from now on we will denote the fields in the 
target ($A^+=0$) gauge by capital letters and fields in the projectile 
($a^-=0$) gauge by lower case letters. This we do for both the 
background part of the field and for the small fluctuation part.  The 
field dependent matrix $V$ is given by 
\begin{equation} 
  V=P\exp\bigg[-i\int_{-\infty}^{x^-}dx^-(b^+ +g a^+)\bigg]\, . 
\end{equation} 
The condition $A^+=0$ does not by itself specify the lower limit of 
the integration over $x^-$ in the exponential.  However, choosing this 
limit to be at minus infinity ensures that $V(x^-\to-\infty)={\sf 1}$ 
and, as a consequence, ${A^i(x^-\to -\infty)=a^i(x^-\to -\infty)}$. 
The projectile gauge fields satisfy the standard vanishing boundary 
conditions at infinity.  This choice of the lower limit of the 
integration, therefore, guarantees that the target gauge fields also 
vanish at $x^-\to-\infty$ and, further, satisfy the residual gauge 
condition ${\partial^i A^i(x^-\to-\infty)=0}$ that was imposed in 
Refs.~\cite{Jalilian-Marian:1997jx,Jalilian-Marian:1997gr,Jalilian-Marian:1997dw,Kovner:1999bj}.  To calculate $\chi$ we 
only need to consider the linearized relation between the small 
fluctuations of the fields in the two gauges. To do this we need to 
expand $V$ to first order in $a^+$. This has been done in the previous 
section. Taking only linear terms in $a^+$ in Eq.~(\ref{vexp}) and 
substituting them into Eq.~(\ref{trans}) we find for the transverse 
components of the field 
\begin{align} 
  A^i_a(x) & = \theta(-x^-)\Big[a^i_a(x)- 
  \int_{-\infty}^{x^-}dx^-\partial^ia^+\Big] + \nonumber\\ 
  & \qquad\qquad +\theta(x^-)\Big[\tilde U^{ab}a^i_b(x)-D^i_{ab}\big( 
  \int_{-\infty}^{0}dx^-a^+_b 
  + \tilde U_{bc}\int_{0}^{x^-}dx^-a^+_c\big)\Big] \nonumber\\ 
  & = \Big[\delta^{ij}-W\partial^iW^\dagger 
  \frac{1}{\partial^+\partial^-} W\partial^j W^\dagger\Big]_{ab} (W 
  a^j)_b\, . 
\label{trans1} 
\end{align} 
Here, the matrix $\tilde U$ is the same as in the previous section and 
is related to the classical background by 
\begin{align} 
  & \tilde U(x_\perp)={\cal P}  
  \exp \{-i\int_{-\infty}^{+\infty} dx^- b^+\}\, ,  
  \nonumber \\ 
  & B^i=\theta(x^-)i\tilde U\partial^i\tilde U^\dagger 
\end{align} 
as per Eq.~(\ref{back}).  We have also defined 
\begin{equation} 
  W(x^-)={\cal P}\exp \bigg\{- i\int_{-\infty}^{x^-}dx^-b^+(x^-)\bigg\} 
  =\theta(-x^-)1+\theta(x^-)\tilde U\, , 
\end{equation} 
which is essentially the classical part of $V$. The operator 
$\frac{1}{\partial^+}$ in the last line of Eq.  (\ref{trans1}) is 
defined as the integral from $-\infty$\footnote{ Eq.~(\ref{trans1}) 
  has been derived also in Refs.~\cite{ayala:1995kg,ayala:1996hx}. The only 
  difference in our derivation is that the meaning of the $\frac{1}{ 
    p^\pm}$ pole is entirely unambiguous and, as discussed above, is 
  dictated by the residual gauge condition.}.  We will further 
simplify this expression by using the on-shellness conditions 
\begin{align} 
  & a^+=\frac{\partial^i}{ \partial^-}a^i\, , \nonumber\\ 
  & (2\partial^-D^+[b^+]-\partial_\perp^2)a^i=0\, . 
\label{onsh} 
\end{align} 
The resulting relation between the on-shell transverse fields in the 
two gauges is 
\begin{multline} 
\label{transf} 
  A^i_a = \theta(-x^-)(t-l)^{ij}a^j_a(x^-) + \\ 
  + \theta(x^-)\Big\{\tilde U_{ab}(t-l)^{ij}a^j_b(x^-)-2D^i_{ab} 
  \Big[\frac{\partial^j}{ \partial_\perp^2}a^j_b(x^-\to 0^-) -\tilde 
  U_{bc}\frac{\partial^j}{ \partial_\perp^2}a^j(x^-\to 
  0^+)\Big]\Big\}\, . 
\end{multline} 
Note that we have to specify on which side of $x^-=0$ the fields are 
taken since the solutions of the small fluctuation equations in the 
projectile gauge are discontinuous at zero.  Now, recall that $a^i$ 
satisfies, everywhere except at $x^-=0$, the free equations of motion. 
With this in mind we can compare this equation with Eq.~(\ref{solut}). 
We see that Eq.~(\ref{transf}) is indeed precisely of the form 
Eq.~(\ref{solut}) with 
\begin{align} 
  \gamma^i_{+,a} & =-2D^i_{ab} \Big[\frac{\partial^j}{ 
    \partial_\perp^2} a^j_b (x^-\to 0^-) 
  -\tilde U_{bc}\frac{\partial^j}{ \partial_\perp^2}  
  a^j_c(x^-\to 0^+)\Big] 
  \, , \nonumber\\ 
  v^i_{-,a} & =(t-l)^{ij} a^j_a(x^-\to 0^-)\, , \nonumber\\ 
  v^i_{+,a} &= \tilde U_{ab} (t-l)^{ij} a^j_b (x^-\to 0^+)\, . 
\label{rels} 
\end{align} 
Remembering that (see for example Ref.~\cite{Hebecker:1998kv}) 
\begin{equation} 
  a^i(x^-\to 0^+)=\tilde U^\dagger a^i(x^-\to 0^-)\, , 
\end{equation} 
we see that the functions $v^i_+, v^i_-$ and $\gamma^i_+$ are related 
precisely by the same relations as in Eq.~(\ref{gammav}). 
 
We have established, therefore, that if $a^i$ satisfies the equation 
of motion in the projectile gauge, then the transformed field $A^i$ of 
Eq.~(\ref{trans}) satisfies the equations of motion in the target 
gauge. The only remaining question is that of the normalization of the 
eigenfunctions.  Recall that the functions $a^i$ in the calculation of 
Ref.~\cite{Balitskii:1996ub}, which was reproduced in the previous section, 
were normalized in the same way as the eigenfunctions of the free 
theory.  That is to say, the full set of on-shell eigenfunctions is 
obtained by choosing $a^i_{p_\perp}(x^-\to 0^-)$ as a complete set of 
normalized eigenfunctions of the unit operator in the transverse space 
\begin{equation} 
  \int\frac{d^2p_\perp}{ 4\pi^2}a^i_{p_\perp}(x_\perp,x^-\to 0^-)  
  a^{j*}_{p_\perp}(y_\perp,x^-\to 
  0^-)=\delta^{ij}\delta(x_\perp-y_\perp)\, . 
\end{equation} 
Since $t-l$ is a unitary operator, Eq.~(\ref{rels}) tells us that 
$v^i_-$ is also normalized to unity rather than to a nontrivial 
operator $O$ as in the JKLW calculation Eq.~(\ref{ort}).  Using this, 
as well as the relations Eq.~(\ref{rels}) and Eq.~(\ref{defchi}), we 
find that when translated into the language of JKLW, the results of 
Ref.~\cite{Balitskii:1996ub} give Eq.~(\ref{chif1}) as the mean fluctuation 
of the induced chromoelectric field.  The essential nonlinearity of 
Eq.~(\ref{chif}) is, therefore, absent in this calculation. 
 
So far, we have only considered the real part of the JKLW kernel, 
$\chi$. Of course, the same method can be applied to find what is the 
form of the virtual part $\sigma$, Eq.~(\ref{final}), that arises from 
the calculation of Ref.~\cite{Balitskii:1996ub}.  To reproduce the virtual 
part it is clearly necessary to keep the quadratic terms in the 
relation between $A^i$ and $a^i$.  Thus the quadratic terms in 
Eq.~(\ref{vexp}) will be important in this calculation.  Other than 
that, the calculation is straightforward.  Again the gauge invariance 
ensures that all the ``kinematical'' factors of $\sigma$ of 
Refs.~\cite{Jalilian-Marian:1997jx,Jalilian-Marian:1997gr,Jalilian-Marian:1997dw,Kovner:1999bj} are reproduced in the 
projectile gauge calculation and the only difference comes from the 
difference in the normalization of the eigenfunctions.  It is clear, 
therefore, that the result of such a calculation is again the lowest 
order expansion of $\sigma$ in powers of $\gamma_+$ 
\begin{multline} 
  {\tilde\sigma}^{i}_{a} = {\Big[\frac{D^i}{D^2}\Big]}_{ab} 
  \Bigg(\frac{N_c}{2} (\partial^j \alpha^j_b) \langle 
  x_\perp|\frac{1}{\partial^2}|x_\perp\rangle \\ 
  - f_{bcd} \langle x_\perp|\Big[ 4 D^j \frac{1}{\partial^2} \partial 
  D \frac{D^j}{D^2} + 2\frac{1}{\partial^2}\partial\alpha - 2 
  \alpha\partial\frac{1}{\partial^2} + 4 \alpha^j \frac{1}{\partial^2} 
  \alpha^j 
  \Big]_{cd} |x_\perp\rangle  \Bigg) \\ 
  - 2 \epsilon^{ij} {\Big[\frac{D^i}{D^2}\Big]}_{ab}(x,y) f_{bcd} 
  \epsilon^{kl} \langle y_\perp|\Big[ D^k \Big( \frac{1}{\partial^2} + 
  \frac{1}{D^2} - \frac{1}{\partial^2} \partial D \frac{1}{D^2} - 
  \frac{1}{D^2} D \partial \frac{1}{\partial^2} \Big) 
  D^l\Big]_{cd}|y_\perp\rangle \, . 
\end{multline} 
 
\subsection{The Doubly Logarithmic Limit.}

Before exploring the relationship between the two approaches further 
in the next section, we want to make a comment about the form 
Eq.~(\ref{chif1}).  Although this equation certainly gives 
$\tilde\chi$ in general as a nonlinear function of the background 
field $\alpha^i$, this nonlinearity disappears in the double 
logarithmic limit.  Following Ref.~\cite{Jalilian-Marian:1998cb} we take the double 
logarithmic limit as the limit when the background field $\alpha^i$ 
does not depend on $x_\perp$.  In this limit, the covariant and the 
simple derivatives commute and it is easy to see that 
Eq.~(\ref{chif1}) reduces to 
\begin{equation} 
  \tilde\chi^{ij}=4\frac{\alpha^2}{ \partial^2}\frac{D^iD^j}{ D^2} 
\end{equation} 
or 
\begin{equation} 
  \tr\tilde\chi=4\tr\frac{\alpha^2}{ \partial^2}\, . 
\end{equation} 
When substituted into the evolution equation Eq.~(\ref{correlf}) this 
gives the simple linear double logarithmic DGLAP evolution for the 
gluon distribution function $G\propto\tr\,\alpha^2$ (see 
Ref.~\cite{Jalilian-Marian:1998cb} for a detailed derivation).  This is in contrast with 
the situation discussed in Ref.~\cite{Jalilian-Marian:1998cb} where the double 
logarithmic limit of Eq.~(\ref{chif}) was studied. It was shown there 
that the nonlinearities in Eq.~(\ref{chif}) survive in the doubly 
logarithmic limit and, in fact, lead even in this limit to the 
``almost saturation'' of the gluon distribution. 
 
The absence of the nonlinearities is in contradiction with the 
explicit calculation of Mueller and Qiu \cite{Mueller:1986wy} who showed that the 
QCD evolution of the gluon distribution in the doubly logarithmic 
approximation does indeed contain contributions from higher twist 
operators.  This again underscores our observation that the 
nonlinearities included in the evolution of Ref.~\cite{Balitskii:1996ub} are 
not the whole story. Those are the ``kinematical'' nonlinearities in 
the sense discussed in the previous section and do not include 
interesting dynamical effects which come into play when the parton 
density becomes large. 
 
In fact, the triviality of the doubly logarithmic limit of the 
calculation of the previous section is easy to understand using the 
intuition of the dipole model approach. In the projectile evolution 
picture, the doubly logarithmic limit is achieved by assuming that, in 
every step in the evolution, the extra gluon that is emitted into the 
virtual photon wave function has the smallest transverse momentum or, 
in the coordinate space, has the largest transverse coordinate 
\cite{Kovchegov:1999yj}.  The part of the wave function that contains this gluon, 
therefore, essentially describes one adjoint dipole of large 
transverse size. One leg of this dipole is the newly emitted gluon, 
while the other leg is the remainder partons which are closely bunched 
together in the coordinate space\footnote{These partons are in the 
  adjoint representation of the color group, since together with the 
  extra emitted gluon the state must be an overall singlet.}. The 
cross section for the scattering of the adjoint dipole in the large 
$N_c$ limit is simply related to the cross section for the fundamental 
dipole $\sigma_{\mathrm{adj}} = 2 \sigma_{\mathrm{fund}} - 
\sigma_{\mathrm{fund}}^2$.  Thus, if initially one starts (like in 
Ref.~\cite{Balitskii:1996ub} or Ref.~\cite{Kovchegov:1999yj}) from a fundamental dipole, 
the appearance of the large adjoint dipole in the wave function leads 
to a nonlinear GLR type quadratic term \cite{Gribov:1984tu} in the evolution 
equation for the scattering cross section. This result indeed has been 
derived in Ref.~\cite{Kovchegov:1999yj}.  However, if we want to consider the 
evolution of the gluon distribution itself, the 
initial state should contain an adjoint rather than a fundamental 
dipole. This can be achieved by 
considering ``DIS'' of a virtual particle that couples to $\tr F^2$ 
\cite{kovchegov:1998bi}. In this case, in any step in the doubly 
logarithmic evolution, the state contains only one adjoint dipole. The 
probability for the appearance of a larger dipole in the approximation 
of independent emissions is itself proportional to the number of 
gluons. The evolution of the gluon distribution in this approximation 
is, therefore, necessarily linear and is merely the simple DGLAP DLA.

\section{More on the target versus projectile gauge} 
\label{sect:more} 
 
The discussion of the previous section may seem a little paradoxical 
on the purely technical level.  Indeed, we have been calculating the 
same physical quantity in two different ways. The quantity in question 
is the equal (lightcone) time propagator of the transverse components 
of the vector potential $A^i$ in the target lightcone gauge. The first 
way of performing the calculation is to work entirely in the target 
gauge as was done in Refs.~\cite{Jalilian-Marian:1997jx,Jalilian-Marian:1997gr,Jalilian-Marian:1997dw,Kovner:1999bj}. This 
gives the result Eq.~(\ref{chif}). The second way to calculate the 
same quantity is to first calculate the propagator of $a^i$ in the 
projectile gauge and then gauge transform the result into the target 
gauge using Eqs.~(\ref{trans1},\ref{transf}). This results in an 
inequivalent expression Eq.~(\ref{chif1}). 
 
\subsection{The $i\epsilon$ complication.}
Our first aim in this section is to resolve this technical paradox. To 
do this let us consider in more detail the calculation of 
Refs.~\cite{Jalilian-Marian:1997jx,Jalilian-Marian:1997gr,Jalilian-Marian:1997dw,Kovner:1999bj} and its transformation 
into the projectile gauge.  The equal time propagator of the 
transverse components of vector potential is calculated in the 
following way \cite{Jalilian-Marian:1997jx,Jalilian-Marian:1997gr,Jalilian-Marian:1997dw,Kovner:1999bj}.  One starts with 
the quadratic part of the action for the small fluctuations of 
$A^\mu$.  Integrating $A^-$ it is reduced to a quadratic action for 
the small fluctuations of the transverse components of the vector 
potential 
\begin{equation} 
  S=\int d^4xd^4y\, A^i(x)G^{-1ij}(x,y)A^j(y)\, . 
\end{equation} 
For the purpose of this discussion we use somewhat simplified 
notations and omit the color indices on the fields. The explicit form 
of $G^{-1}$ is given in Refs.~\cite{Jalilian-Marian:1997jx,Jalilian-Marian:1997gr,Jalilian-Marian:1997dw,Kovner:1999bj}. 
One then finds properly normalized eigenfunctions of $G^{-1}$ 
\begin{align} 
  & G^{-1ij}(x,y)A^j_{\lambda,p^-,r}(y)=\lambda A^i_{\lambda,p^-,r}(x) 
  \, , \nonumber\\ 
  &\int d^4xA^i_{\lambda,p^-,r}(x)A^{i*}_{\lambda',p^{-'},r'}(x)= 
  \delta(\lambda-\lambda')\delta(p^--p^{-'})\delta^2(r-r')\, . 
\end{align} 
Using the complete set of eigenfunctions one constructs the propagator 
with the standard $i\epsilon$ prescription as 
\begin{equation} 
  G^{ij}(x,y)=\int \frac{d \lambda}{ \lambda+i\epsilon} 
  \int d p^- d^2r  
  A^i_{\lambda,p^-,r}(x)A^{j*}_{\lambda,p^{-},r}(y)\, . 
\label{g} 
\end{equation} 
The limit $x^+=y^+$, and $x^-,y^-\to\infty$ is then taken to calculate 
$\chi^{ij}$.  Clearly, the equal time limit selects the on-shell 
eigenfunctions $\lambda=0$ and, therefore, when transforming into the 
projectile gauge it is important to keep track of the $i\epsilon$ 
prescription. The simplest way to do this is to include the 
$i\epsilon$ term directly in the action 
\begin{equation} 
  S_t=\int d^4xd^4yA^i(x)\Big[  
  G^{-1ij}(x,y) +i\epsilon\delta^{ij}\delta(x-y)\Big]A^j(y)\, . 
\label{sq} 
\end{equation} 
The propagator Eq.~(\ref{g}) is then just the inverse of the quadratic 
form in Eq.~(\ref{sq}) without any additional regulators. 
 
To transform this expression into the projectile gauge one has to use 
Eqs.~(\ref{trans1},\ref{transf}). The gauge invariance of the QCD 
action ensures that the first term in Eq.~(\ref{sq}) under this 
transformation transforms into 
\begin{equation} 
  a^i(x)D^2(x,y)a^i(y)\, , 
\end{equation} 
which is exactly the action used in Sec.~\ref{sect:bk} to calculate 
the projectile gauge propagator. However, the $i\epsilon$ term is not 
so simple.  If the transformation Eq.~(\ref{transf}) was unitary, the 
norm of the field $A^i$ would be preserved and the $i\epsilon$ term in 
Eq.~(\ref{sq}) would transform into the standard $i\epsilon \int 
d^4xa_i(x)a_i(x)$ term in the projectile gauge.  The problem is that, 
as we saw in the previous section, the transformation 
Eq.~(\ref{transf}) is not unitary. A normalized function $a^i$ is 
transformed into a function $A^i$ normalized not to unity but rather 
to an eigenvalue of the operator $O$ in Eq.~(\ref{o}).\footnote{The 
  fact that the transformation between the two gauges is non-unitary 
  is not unusual.  Even though it is a gauge transformation and 
  therefore formally unitary, the gauge parameter itself depends on 
  the dynamical field.  Such transformations are generically 
  non-unitary and do not preserve the scalar product.}  This was 
precisely the root for the discrepancy between $\chi$ and 
$\tilde\chi$.  The resulting projectile gauge action can be written 
as\footnote{In writing this expression we have made use of the fact 
  that the $i\epsilon$ term is important only for functions $a^i$ that 
  satisfy $D^2a^i=0$.} 
\begin{multline} 
  S_p = \int d^4xa^i(x)D^2a^i(x) + \\ 
  +i\epsilon \int d^4x \Big[a^i(x)a^i(x)+2 
  a^i(0^-)[\partial^i\frac{1}{\partial_\perp^2} -D^i\frac{1}{ 
    D_\perp^2}]D^2_\perp [\partial^i\frac{1}{\partial_\perp^2} 
  -D^i\frac{1}{ D_\perp^2}]a^i(0^-)\Big]\, . 
\end{multline} 
Thus, the standard $i\epsilon$ prescription in the target gauge is 
equivalent to a fairly complicated momentum dependent prescription in 
the projectile gauge.  Since the calculation of Sec.~\ref{sect:bk}, 
following Ref.~\cite{Balitskii:1996ub}, was performed using the standard 
$i\epsilon$ prescription in the projectile gauge the result is, 
indeed, expected to differ from that of 
Refs.~\cite{Jalilian-Marian:1997jx,Jalilian-Marian:1997gr,Jalilian-Marian:1997dw,Kovner:1999bj}. 
 
While the technical reason for the difference between the results of 
Refs.~\cite{Jalilian-Marian:1997jx,Jalilian-Marian:1997gr,Jalilian-Marian:1997dw,Kovner:1999bj} and Ref.~\cite{Balitskii:1996ub} 
is clear, the physics behind it is not so obvious. In the rest of this 
section, we will make an attempt to understand the physical reason for 
this difference. 
 
As we have just explained, the calculation of 
Refs.~\cite{Jalilian-Marian:1997jx,Jalilian-Marian:1997gr,Jalilian-Marian:1997dw,Kovner:1999bj} is equivalent to a 
calculation in the projectile gauge with a nonstandard momentum 
dependent $i\epsilon$ prescription. It is well known that such a 
change of prescription is equivalent to a calculation not in the 
vacuum state but rather in a state which contains 
gluons \cite{Hoyer:1996wd, Hoyer:1999mn, Hoyer:1999mn}. We, therefore, ask ourselves why 
the projectile gauge calculation should be performed in a state which, 
on top of the background field $b^+$, also contains additional gluons. 

\subsection{Evolution as Renormalization Group in the projectile gauge.}
To answer this question let us first try to reformulate the projectile 
gauge calculation of Sec.~\ref{sect:bk} in terms of the Wilson 
renormalization group akin to the approach of 
Refs.~\cite{Jalilian-Marian:1997jx,Jalilian-Marian:1997gr,Jalilian-Marian:1997dw,Kovner:1999bj}.  The hadron is 
represented as a statistical ensemble of the static $b^+$ fields of 
the form Eq.~(\ref{aplus}) with a statistical weight $Z[b]$. 
Evolution in $x$ generates induced vector potential which changes the 
statistical weight.  Strictly speaking the induced vector potential is 
not static.  It has components in the frequency range 
$p^-<\Lambda\propto 1/x$.  However, as long as the frequency of the 
components of the projectile wave function are large enough, one can 
treat the induced potential as static during the interaction with the 
$q\bar q$ pair.  Also, as long as the wavelength of the projectile in 
the $x^-$ direction is large enough ($p^+$ is small) the induced 
vector potential can be approximated by a $\delta(x^-)$ shaped 
function.  One can, equivalently, describe the hadron by a statistical 
ensemble of $V$ and $V^\dagger$ with some statistical weight 
$Z[V,V^\dagger]$.  It is clear that both descriptions encode exactly 
the same information. For this purpose one has to define analogs of 
$\chi$ and $\sigma$, that is the (connected) fluctuation correlation 
functions of order $\alpha_s$. The resulting evolution equation 
\begin{multline} 
\label{final+} 
  \frac{d}{d\ln \frac{1}{x}}Z[U,U^\dagger]= \\ 
  \alpha_s \Bigg[ \frac{1}{2} \Big( \frac{\delta^2}{\delta U(u)\delta 
    U^\dagger (v)}\big[Z {\chi}_{q{\bar q}} (u, v) \big] + 
  \frac{\delta^2}{\delta U^\dagger (u)\delta U(v)} 
  \big[Z {\chi}_{{\bar q}q} (u, v) \big] \\ 
  + \frac{\delta^2}{\delta U(u)\delta U(v)}\big[Z {\chi}_{qq} (u, v) 
  \big] 
  + \frac{\delta^2}{\delta U^\dagger (u)\delta U^\dagger (v)}  
  \big[Z {\chi}_{{\bar q}{\bar q}} (u, v) \big] \Big) \\ 
  - \frac{\delta}{\delta U(u)}\big[Z{\sigma}_{q} (u) \big] - 
  \frac{\delta}{\delta U^\dagger(u)}\big[Z{\sigma}_{{\bar q}} (u) 
  \big] \Bigg] 
\end{multline} 
is the analog of Eq.~(\ref{final}).

Using the formulae of Sec.~(\ref{sect:bk}) and the appendix we find 
\begin{equation} 
  {\sigma}_{q}^{\alpha\beta} (x_\perp ) = 
  -\frac{1}{2\pi^2}\int d^2 z_\perp \frac{1}{(x - 
    z)^2_\perp}  
  \Big[ \tr{\big( U(x_\perp )\cdot U^\dagger (z_\perp)\big)} {\big( 
    U(z_\perp )\big)}^{\alpha\beta}  
  - N_c {\big( U(x_\perp )\big)}^{\alpha\beta} \Big]\, , 
\end{equation} 
 
\begin{align} 
  {\chi}_{q{\bar q}}^{\alpha\beta , \gamma\delta} (x_\perp , y_\perp ) 
  & =  \frac{1}{2\pi^2} \int d^2 z_\perp 
  \frac{(x - z)_\perp \cdot (y - z)_\perp }{(x - z)^2_\perp (y - z)^2_\perp } 
\nonumber \\ 
  &\Big[ {\big( U(z_\perp ) \cdot U^\dagger 
    (y_\perp)\big)}^{\alpha\delta} 
  {\big( U^\dagger (z_\perp ) \cdot  U (x_\perp)\big)}^{\gamma\beta} 
\nonumber \\ 
  &+ {\big( U(x_\perp ) \cdot U^\dagger (z_\perp)\big)}^{\alpha\delta} 
  {\big( U^\dagger (y_\perp ) \cdot  U (z_\perp)\big)}^{\gamma\beta} 
\nonumber \\ 
  &- {\delta}^{\alpha\delta} {\big( U^\dagger (y_\perp ) \cdot 
    U(x_\perp)\big)}^{\gamma\beta} - {\big( U (x_\perp) \cdot 
    U^\dagger (y_\perp ) \big)}^{\alpha\delta} {\delta}^{\gamma\beta} 
  \Big]\, , 
\end{align} 
 
\begin{align} 
  {\chi}_{qq}^{\alpha\beta , \gamma\delta} (x_\perp , y_\perp ) & = 
  -\frac{1}{2\pi^2}\int d^2 z_\perp 
  \frac{(x - z)_\perp \cdot (y - z)_\perp }{(x - z)^2_\perp (y - z)^2_\perp } \nonumber \\ 
  &\Big[ {\big( U(z_\perp ) \big)}^{\alpha\delta} {\big( U (y_\perp) 
    \cdot U^\dagger (z_\perp ) \cdot U (x_\perp)\big)}^{\gamma\beta} 
\nonumber \\ 
  &+ 
  {\big(U (x_\perp) \cdot U^\dagger (z_\perp ) \cdot U (y_\perp) 
    \big)}^{\alpha\delta} 
  {\big( U(z_\perp ) \big)}^{\gamma\beta} \nonumber \\ 
  & - {\big( U(x_\perp ) \big)}^{\alpha\delta} {\big( U(y_\perp ) 
    \big)}^{\gamma\beta} - {\big( U(y_\perp ) \big)}^{\alpha\delta} 
  {\big( U(x_\perp ) \big)}^{\gamma\beta} \Big] 
\label{chinn}
\end{align} 
${\sigma}_{\bar q}$ is obtained from ${\sigma}_q$ by replacing $U$s by 
$U^\dagger$s. The same is true for ${\chi}_{{\bar q}{\bar q}}$ and 
${\chi}_{qq}$.  ${\chi}_{{\bar q}q}$ is obtained from ${\chi}_{qq}$ by 
swapping $(x,\alpha\beta)$ with $(y,\gamma\delta)$ as this exchanges 
the $q$ and $\bar q$ lines there. 
We stress that Eqs.(\ref{final+}-\ref{chinn}) contain all 
the information that is contained in the BK equation as well as in the equations for higher correlation functions of $U$ that appear in Ref.\cite{Balitskii:1996ub}.

\subsection{The snag.}
There is one implicit assumption in this procedure: namely that the $a^+$ 
component of the vector potential is the only relevant one. One assumes that 
if, for 
example, an $a^i$ component is generated in the evolution it does not 
affect the subsequent evolution of the physical cross section.  This 
is, however, not quite right. What is important for the interaction 
with the projectile is not merely $b^+$ but rather the $F^{+i}$ 
component of the color electric field.  The interaction between the 
projectile and the target is due to the term $F^{-i}F^{+i}$ in the QCD 
Lagrangian. The $F^{-i}$ component is the Weisz\"aker-Williams field 
of the $q\bar q$ pair, while the $F^{+i}$ component is generated by 
the color charges in the target. In the eikonal approximation it is 
true that $F^{+i}=\partial^i b^+$ and, therefore, the coupling can be 
written as $b^+J^-$, where $J^-=\partial^iF^{-i}$.  However, if there 
is a contribution to $F^{+i}$ coming from the transverse component of 
the vector potential, it should be taken into account. 
 
It is easy to see that such a contribution is indeed generated by the 
low $x$ evolution.  Suppose one starts the evolution initially with 
the background field configuration as in Eq.~(\ref{aplus}). In the 
first step of the evolution one generates both the increment in $a^+$ 
and the increment in $a^i$. The two are related by the condition 
Eq.~(\ref{onsh}) 
\begin{equation} 
  a^+=\frac{\partial^i}{ \partial^-}a^i\, . 
\end{equation} 
Naively, one would expect that since all the fluctuation fields in 
this step have small frequencies, it should be true that $a^+\ll a^i$ 
and, therefore, it should be safe to forget about $a^i$. The reason 
this is incorrect is that the on-shell solutions for $a^i$ are 
discontinuous at $x^-=0$. Therefore, even though the field $a^i$ is 
indeed small, it has a large derivative with respect to $x^-$ which 
contributes to the field strength.  In fact, the induced 
chromoelectric field is 
\begin{equation} 
  \delta F^{+i}=\partial^i a^+-D^+a^i\, . 
\end{equation} 
Recalling that on-shell $a^i$ satisfy the second of the equations 
Eq.~(\ref{onsh}), we see that the second term in this expression is 
$\frac{\partial^2_\perp}{ 2\partial^-}a^i$ and is of the same order as 
the first term $\frac{\partial^i\partial^j}{ \partial^-}a^j$. Clearly, 
even if one starts initially from a background which only contains 
$b^+$ after long enough evolution a large transverse component of the 
vector potential is generated.  When the contribution of the 
transverse component to the field strength is comparable to the 
contribution of the ``$+$'' component, the eikonal approximation 
breaks down and the evolution discussed in Sec.~\ref{sect:bk} ceases 
to be valid.  It looks indeed very natural that in order to take into 
account the presence of the (potentially large) transverse field, the 
calculation in the projectile gauge should be performed around a state 
that contains transverse gluons apart from the $b^+$ background. 
 
One could try to argue that the transverse part of the vector 
potential can be somehow gauged away and the calculation could still 
be performed consistently around a pure $b^+$ background. Even if this 
is possible the evolution of the background defined by such a 
procedure will be different from the evolution of Sec.~\ref{sect:bk}. 
In any case, we do not see how such ``regauging'' is possible.  

It is 
instructive to see in more detail how the gauge fixing works in both 
the projectile and the target gauges and why the two seem to have 
different status as far as the renormalization group structure is 
concerned.  As we mentioned above, the chromoelectric field is created 
by the color charges in the target.  In fact, the whole 
renormalization group procedure can be formulated in terms of the 
color charge density $j^+$ rather than the vector potentials 
themselves, which was in fact originally done in 
Refs.~\cite{Jalilian-Marian:1997jx,Jalilian-Marian:1997gr,Jalilian-Marian:1997dw}.  The background vector 
potentials are found as static solutions of classical equations of 
motion in the presence of the color charge density 
$j^+=\rho\delta(x^-)$ 
\begin{align} 
  & F^{ij}=0\, ,\nonumber\\ 
  & D^i F^{+i}=D^i[D^ib^+-\partial^+b^i]=j^+\, . 
\end{align} 
An important property of these equations is that for a given $\rho$ 
they have infinitely many solutions. By considering an arbitrary 
unitary matrix $V(x_\perp,x^-)$ it is straightforward to see that all 
the following are solutions 
\begin{align} 
  &b^i=iV^\dagger\partial^i V \, ,\nonumber\\ 
  & b^+=\frac{1}{ D^2_\perp}[j^++D^i\partial^+b^i]\, . 
\label{sols} 
\end{align} 
 
The difference between the target and the projectile gauges at this 
point becomes important.  In the target gauge $B^+=0$, the equations 
reduce to 
\begin{align} 
  & B^i=iV^\dagger\partial^i V\nonumber\, ,\\ 
  & D^i\partial^+B^i=-j^+\, . 
\end{align} 
We, therefore, get rid of almost all the solutions, the only residual 
degeneracy being the value of the matrix $V$ at $x^-\to-\infty$.  The 
imposition of the residual gauge condition ${\partial^i 
  A^i(x^-\to-\infty)}$ then removes all solutions except one.  In the 
projectile gauge the situation is very different.  The condition 
$a^-=0$ does not eliminate any of the infinite number of solutions 
Eq.~(\ref{sols}). The choice of the residual gauge fixing is thus 
crucial to eliminate the redundant solutions. If those are not 
eliminated, the perturbative calculation will be plagued with zero 
mode problems.  The calculation in Sec.~\ref{sect:bk} was in fact 
performed with the residual gauge fixing $\partial^i a^i(p^-=0)=0$. 
This gauge fixing does indeed eliminate all the solutions except the 
one which has vanishing $a^i$ and has therefore precisely the form of 
Eq.~(\ref{aplus}). 
 
Now, consider the renormalization group calculation. Here, we have to 
integrate out modes which have higher frequency $p^-$. In the target 
gauge this is straightforward: the residual gauge condition does not 
care about frequency. It therefore eliminates nonzero frequency 
fluctuation modes which do not vanish at $x^-\to-\infty$ in the same 
way as it eliminated the static background solutions with this 
behavior. As a result, the fluctuation modes have a very similar 
structure to the background field and the induced field is similar to 
the background.  It is, therefore, straightforward to formulate a 
selfsimilar renormalization group transformation in this gauge.  The 
situation is quite different in the projectile gauge. The residual 
gauge condition, although it fixes unambiguously the background, has 
nothing to say about the fluctuations --- it only fixes the static 
modes! It is impossible therefore to ensure that the fluctuations will 
have the same form as the static background.  In fact, as we have seen 
above, it will not be the case. The first equation of 
Eq.~(\ref{onsh}) in the projectile gauge is just one of the equations 
of motion (with or without the external source).  This means that it 
is possible to have nonvanishing $a^+$ with vanishing $a^i$ only at 
exactly zero frequency. At any finite frequency nonvanishing, $a^i$ is 
required.  As we have seen, this $a^i$ contributes to the induced 
chromoelectric field, or equivalently to the induced color charge 
density. The calculation thus explicitly lacks a selfsimilar structure 
and proper renormalization group setup does not seem 
possible\footnote{It may be possible to reformulate RG so that it 
  would include also transverse background fields or equivalently 
  finite number of gluons in addition to $b^+$. This seems, however, 
  to be quite a complicated problem and is far beyond the scope of our 
  discussion here.} unless extra eikonal approximation is invoked. 
 
The discussion of this section leads us to conclude that the 
projectile gauge calculation, as formulated in Ref.~\cite{Balitskii:1996ub} 
and Sec.~\ref{sect:bk} is only valid as long as the eikonal 
approximation is applicable. When the evolution is continued for a 
large span of $1/x$, the eikonal approximation breaks down and the 
higher nonlinear corrections of 
Refs.~\cite{Jalilian-Marian:1997jx,Jalilian-Marian:1997gr,Jalilian-Marian:1997dw,Kovner:1999bj} should become important. 
This is not to say that one cannot learn much from this simplified 
evolution.  Quite to the contrary --- clearly there is a range of $x$ 
values where this evolution captures the relevant physics. This is 
particularly true when the target is large --- the case of a large 
nucleus discussed in Ref.~\cite{Kovchegov:1999yj}. In this case, the eikonal cross 
section is significantly different from the simple perturbative one 
which assumes single scattering.  The nonlinearity of the evolution 
becomes important much faster than for a small hadron. One, therefore, 
expects unitarization to appear already within the eikonal regime. 
Subsequent appearance of other nonlinear corrections will not change 
the fact that the total cross section has unitarized. It must, 
however, affect other more exclusive properties of the process such as 
the structure of final states.  The spectrum of the target gauge 
fields is presumably directly related to the spectrum of the emitted 
gluons \cite{kovchegov:1998bi,Mueller:1999fp}.  Thus, when the evolution of 
these fields changes even locally one expects this change to be 
visible in the spectrum of final state gluons. Assuming the local 
parton-hadron duality this then has to be mirrored in the spectrum of 
final state hadrons.

\section{Conclusions} 
\label{sect:conc} 
 
In recent years several approaches to the evolution of dense gluonic 
systems in the saturation regime were developed.  The approaches 
differ from one another in many technical respects and relationship 
between the physics is also not always clear.  In this paper, our aim 
was to relate two of these approaches and thereby to try and reduce 
the entropy in the field.  We have shown that the nonlinear JKLW 
equation of Refs.~\cite{Jalilian-Marian:1997jx,Jalilian-Marian:1997gr,Jalilian-Marian:1997dw,Kovner:1999bj} coincides with 
the BK evolution equation derived in Ref.~\cite{Balitskii:1996ub} and 
\cite{Kovchegov:1999yj} as long as the gluon field induced by the evolution is 
small. We have argued that the approach of Ref.~\cite{Balitskii:1996ub} 
should break down when the field is large enough so that the eikonal 
approximation intrinsic in the derivation of Ref.~\cite{Balitskii:1996ub} 
ceases to be valid. We have also argued that the evolution of 
Refs.~\cite{Jalilian-Marian:1997jx,Jalilian-Marian:1997gr,Jalilian-Marian:1997dw,Kovner:1999bj} when translated into the 
language of Ref.~\cite{Balitskii:1996ub} corresponds to taking into account 
some non-eikonal contributions. 
 
We should note that our discussion puts into perspective the 
discrepancy between the double logarithmic limit (DLL) of the 
evolution of Refs.~\cite{Jalilian-Marian:1997jx,Jalilian-Marian:1997gr,Jalilian-Marian:1997dw,Kovner:1999bj} and the 
evolution suggested in Refs.~\cite{AyalaFilho:1997du,Levin:1998jq,GayDucati:1999sx}.  The AGL equation 
\cite{AyalaFilho:1997du,Levin:1998jq,GayDucati:1999sx} has been shown to arise from the BK equation 
\cite{Kovchegov:1999yj} in the regime where the evolution on the projectile side 
is dominated by production of small size dipoles, or equivalently 
large transverse momentum gluons \cite{Kovchegov:1999yj,Levin:1999mw}. This is a 
natural regime when the target is a small object rather than a large 
object of the typical hadronic size. It was suggested in 
Ref.~\cite{Levin:1999mw} that these configurations also dominate in 
usual DIS in the saturation regime.  It seems to us that this point 
warrants further study. In any case this is not the standard DLL, 
where the evolution on the projectile side is dominated by large 
dipoles.  There is, therefore, no reason to expect that the DLL of 
Refs.~\cite{Jalilian-Marian:1997jx,Jalilian-Marian:1997gr,Jalilian-Marian:1997dw,Kovner:1999bj,Jalilian-Marian:1998cb} has much to do with 
the AGL equation. In fact, as we have shown the DLL of the BK 
evolution is itself extremely simple when considered as the evolution 
of the gluon distribution operator rather than the physical DIS cross 
section. It turns out to be entirely devoid of nonlinear corrections 
and coincides with the standard DGLAP double logarithmic equation. 
The DIS cross section still evolves nonlinearly and in fact saturates 
in this limit due to the nonlinear Glauber type relation between the 
cross section and the gluon distribution \cite{Kovchegov:1999ua}.  On the other 
hand, the DLL of Refs.~\cite{Jalilian-Marian:1997jx,Jalilian-Marian:1997gr,Jalilian-Marian:1997dw,Kovner:1999bj,Jalilian-Marian:1998cb} is 
also nonlinear for the gluon distribution and as a result the 
evolution is slowed down already on the level of the gluon 
distribution\cite{Jalilian-Marian:1999cf}. 
 
We hope that this paper clarifies to some extent the relationship
between the different approaches to the nonlinear low $x$ evolution.
There are still many questions to be answered. In particular, it is
desirable to find a more explicit relation between the nonlinearities
of the JKLW equation and the breakdown of the eikonal approximation
and to better understand the physics of these nonlinearities. Perhaps
the most interesting question concerns the effect of these
nonlinearities on the structure of the final states. Some work on the
analytic understanding of quantities less inclusive than the total
cross section has appeared recently \cite{Kovchegov:1999kx,Kovchegov:1999ji,Kovchegov:1999ep}.
There is also an ongoing numerical effort in connection with the heavy
ion physics \cite{Krasnitz:1998ns,Krasnitz:1999wc} in the framework of the
McLerran-Venugopalan model \cite{mclerran:1994ni,mclerran:1994ka}. Further progress in this
direction is extremely important both for our understanding of the
nonlinear physics and for disentangling linear and nonlinear effects
in the existing data.
 
\vspace{5mm} {\bf Acknowledgments} The work of A.K. is supported by PPARC.
The work of J.G.M. is supported by 
PRAXIS XXI/BD/11277/97 grant (Subprograma Ci\^encia e Tecnologia do 
2$^{\underline o}$ Quadro Comunit\'ario de Apoio --- Portugal). He is 
grateful to the CERN Theory Division, J.~Barbosa, J.~Seixas and 
P.~Sonderegger for hospitality during June-August 1999. H.W. was 
supported by the EC TMR Program, contract ERB FMRX-CT96-0008. He wants 
to thank the members of the Nuclear Theory Group at Brookhaven 
National Lab for their hospitality during October and November 1999. 
\vspace{5mm} 
 
\appendix 
\section{Appendix} 
\label{appendix} 
 
In this appendix, we give a more detailed derivation of the evolution 
equation of Sec.~\ref{sect:bk} including the calculation of the 
contributions where the glue does not interact with the background 
(target). 
 
We start with the quadratic action for small fluctuations in the 
projectile gauge 
\begin{equation} 
  \begin{split} 
    S = \frac{1}{2} \bigg\{& a^{+}_{a} \big[ - ( {\partial}^{-})^{2} 
    \big] a^{+}_{a} 
    - 2 ({\partial}^{i} a^{+}_{a})({\partial}^{-} a^{i}_{a}) \\ 
    & - a^{i}_{a} \Big[ \big( 2 D^{+}_{ab}[b]{\partial}^{-} - 
    ({\partial}_{\perp})^{2} {\delta}_{ab}  \big) 
    {\delta}^{ij} + {\partial}^{i}{\partial}^{j} {\delta}_{ab} \Big] 
    a^{j}_{b} \bigg\}\, . 
\end{split} 
\label{actiona} 
\end{equation} 
The equation of motion for $a^{+}$ is 
\begin{equation} 
  a^{+} = \frac{\partial^i}{\partial^-} a^i\, . 
\end{equation} 
Substituting this in Eq.~(\ref{actiona}) we get 
\begin{equation} 
  S = -\frac{1}{2} a^{i}_{a} (D^{2})_{ab} {\delta}^{ij} a^{j}_{b}\, , 
\end{equation} 
where 
\begin{equation} 
  (D^{2})_{ab} = 2 D^{+}_{ab}[b] {\partial}^{-}  
  - ({\partial}_{\perp})^{2} {\delta}_{ab}  
\end{equation} 
is the same as the inverse propagator for a charged scalar field in 
the presence of a background field $b^{+}$. It has been calculated, 
for example, in Ref.~\cite{Hebecker:1998kv}.  The most useful form for us is 
\begin{multline} 
  - i\left[\frac{1}{(D^{2})}\right]_{ab} = \int\frac{dp^-}{2 p^- 
    (2\pi)^3} 
  \big[\theta (x^- - y^-) \theta (p^-) - \theta (y^- - x^-)\theta (-p^-) \big]  
  \times \\ 
  \times \int d^2 p_{\perp} d^2 q_{\perp} e^{-ip\cdot x+iq\cdot y} 
  \int \frac{d^2 z_{\perp}}{(2\pi)^2} e^{-i( p_{\perp} - q_{\perp}) 
    z_{\perp}} \tilde{U}^{-1}_{ab} (x^-, y^-, z_{\perp}) 
\end{multline} 
with ${p}^{+} =\frac{{p}_{\perp}^{2}}{2p^-}$, ${q}^{+}
=\frac{{q}_{\perp}^{2}}{2p^-}$ and ${q}^{-} = {p}^{-}$.  The color
matrix $ \tilde{U}^{-1}_{ab}(x^-, y^-, z_{\perp})$ is\footnote{More
  rigorously, the structure of $ \tilde{U}^{-1}$ is given by $\tilde
  U_{ab}^{-1}(z_\perp) = e^{i(\theta(x^-) - \theta(y^-)) b(z_\perp)}$.
  However, the difference between this expression and that given in
  Eq.(\ref{tUrep}) only shows up if it is multiplied by $\partial^+$
  derivatives or $\delta(x^-)$ factors. Since we encounter no such
  factors in our calculation we will be using Eq.(\ref{tUrep})
  throughout.}
\begin{multline} 
  \tilde{U}^{-1}_{ab}(x^-, y^-, z_{\perp}) = \big( 
  \theta(-x^-)\theta(-y^-) + \theta(x^-)\theta(y^-)\big) {\delta}_{ab} \\ 
  + \theta(-x^-)\theta(y^-) \tilde{U}_{ab}(z_{\perp}) + 
  \theta(x^-)\theta(-y^-) \tilde{U}_{ab}^{\dagger} (z_{\perp})\, . 
\label{colmat} 
\end{multline}

The on-shell two-point correlator of $a^+$ can be written as 
\begin{multline} 
  \big\langle a_a^+ (x^+ = 0, x_\perp , x^-) a_b^+ (y^+ = 0, y_\perp , 
  y^-)\big\rangle = \big\langle \frac{\partial^i_x}{\partial^-_x} 
  a_a^i (x^+ = 0, x_\perp , x^-) 
  a_b^i (y^+ = 0, y_\perp , y^-) \frac{\partial^i_y}{\partial^-_y}\big\rangle  \\ 
  = - \partial^i_x \partial^i_y {\int} \frac{dp^-}{p^-} \frac{1}{2(2 
    \pi)} \frac{1}{(p^-)^2} 
  \big[\theta (x^- - y^-)\theta (p^-) - \theta (y^- - x^-)\theta (-p^-) \big]  
 \times \\ 
  \int d^2 z_{\perp} \int \frac{d^2 p_{\perp}}{(2\pi)^2} e^{+ i 
    p_{\perp} (x_{\perp} - z_{\perp})} \int \frac{d^2 
    q_{\perp}}{(2\pi)^2} e^{+ i q_{\perp} (z_{\perp} - y_{\perp})} 
  e^{-i \frac{{p_{\perp}}^2}{2 p^-} x^-} e^{+i\frac{{q_{\perp}}^2}{2 
      p^-} y^-} \tilde{U}^{-1}_{ab} (x^-, y^-, z_{\perp})\, . 
\label{correlatora} 
\end{multline}

We now need to expand the eikonal factors 
\begin{equation} 
  V(x_\perp ) = {\cal P} \exp \Big[ -i  
  \int_{-\infty}^{+\infty} dx^- (b^+ +g a^+) (x_\perp , x^-) \Big] 
\end{equation} 
to second order in the fields.  Recalling that the background part of 
the field $b^+ \propto\delta(x^-)$ and that the fluctuation field 
$a^+$ is nonsingular at $x^-=0$, this becomes 
\begin{align} 
  V(x^+ = 0, x_\perp ) &= 
  {\cal P} \exp \Big[ -i g \int_{-\infty}^{0} dx^-  
  a^+ (x^+ = 0, x_\perp , x^-) \Big] \times \nonumber \\ 
  & \qquad \times U(x_\perp ) {\cal P} \exp \Big[ -i g
  \int_{0}^{+\infty} dx^- a^+ (x^+ = 0, x_\perp , x^-) \Big]\, , 
\end{align} 
where 
\begin{equation} 
  U(x_\perp ) = {\cal P} \exp \Big[ -i  
  \int_{-\infty}^{+\infty} dx^- b^+ (x^+ = 0, x_\perp , x^-) \Big] 
\end{equation} 
is the classical part of the eikonal factor. 
 
To second order in $a^+$ we have 
\begin{equation} 
  \begin{split} 
    V(x^+ = 0,& x_\perp ) = \\ & \Big\{ 1 - i g \int_{-\infty}^{0} dx^- 
    a^+ (x^+ = 0, x_\perp , x^-) \\ & \ \  
    - g^2 \int_{-\infty}^{0} dx^- dy^-  
    \theta (y^- - x^-) a^+ (x^+ = 0, x_\perp , x^-)  
    a^+ (x^+ = 0, x_\perp , y^-) \Big\} \\ 
    & \times U(x_\perp )\ \Big\{ 1 - i g \int_{0}^{+\infty} dx^- a^+ 
    (x^+ = 0, x_\perp , x^-) \\& \ \ - g^2 \int_{0}^{+\infty} dx^- dy^- 
    \theta (y^- - x^-) a^+ (x^+ = 0, x_\perp , x^-) a^+ (x^+ = 0, 
    x_\perp , y^-) \Big\}\, , 
  \end{split} 
\end{equation} 
or 
\begin{multline} 
  V(x_\perp ) =  U(x_\perp ) - \\ 
  - i g \Big\{ \int_{-\infty}^{0} dw^- a^+ (x^+ = 0, x_\perp , w^-) U( 
  x_\perp ) 
  + U(x_\perp )  \int_{0}^{+\infty} dw^- a^+ (x^+ = 0, x_\perp , w^-) \Big\} \\ 
  - g^2\Big\{ \int_{-\infty}^{0} dw^- dz^- \theta (z^- - w^-) a^+ (x^+ = 
  0, x_\perp , w^-) a^+ (x^+ = 0, x_\perp , z^-) 
  U(x_\perp ) \\ 
  + U( x_\perp )\int_{0}^{+\infty} dw^- dz^- \theta (z^- - w^-) a^+ 
  (x^+ = 0, x_\perp , w^-) a^+ (x^+ = 0, 
  x_\perp , z^-) \\ 
  + \int_{-\infty}^{0} dw^- a^+ (x^+ = 0, x_\perp , w^-) U( x_\perp ) 
  \int_{0}^{+\infty} dz^- a^+ (x^+ = 0, x_\perp , z^-) \Big\}\, . 
\end{multline} 
Similarly, 
\begin{multline} 
  V^{\dagger} (y_\perp ) =  U^{\dagger} ( y_\perp ) + \\ 
  + i g\Big\{ U^{\dagger} (y_\perp ) \int_{-\infty}^{0} dz^- a^+ (y^+ = 
  0, y_\perp , z^-) 
  + \int_{0}^{+\infty} dz^- a^+ (y^+ = 0, y_\perp , z^-) 
  U^{\dagger} (y_\perp ) \Big\} \\ 
  - g^2\Big\{ U^{\dagger} (y_\perp ) \int_{-\infty}^{0} dw^- dz^- \theta 
  (w^- - z^-) a^+ (y^+ = 0, y_\perp , 
  w^-) a^+ (y^+ = 0, y_\perp , z^-) \\ 
  +\int_{0}^{+\infty} dw^- dz^- \theta (w^- - z^-) a^+ (y^+ = 0, 
  y_\perp , w^-) a^+ (y^+ = 0, y_\perp , z^-) 
  U^{\dagger} (y_\perp ) \\ 
  + \int_{0}^{+\infty} dw^- a^+ (y^+ = 0, y_\perp , w^-) 
  U^{\dagger}(y_\perp ) \int_{-\infty}^{0} dz^- a^+ (y^+ = 0, y_\perp 
  , z^-) \Big\}\, . 
\end{multline} 
 
Rather than directly calculating the eikonal cross section, we will 
first calculate the tensor product of two eikonal factors and later 
take the trace over the color indices.  We use the notation $A\otimes 
B=A^{\alpha\beta}B^{\gamma\delta}$.  To second order in the 
fluctuation field we have 
\begin{equation} 
\begin{split} 
  & {\big\langle V( x_\perp ) \otimes V^\dagger ( y_\perp ) 
    \big\rangle } 
  -  {\big\langle  U( x_\perp ) \otimes U^\dagger ( y_\perp ) \big\rangle }= \\ 
  & \quad = g^2 \big\langle \int_{-\infty}^{0} dw^- a^+ (x^+ = 0, x_\perp 
  , w^-) U( x_\perp ) 
  \otimes U^\dagger (y_\perp ) \int_{-\infty}^{0} dz^- a^+ (y^+ = 0, y_\perp , z^-) \\ 
  & \qquad + \int_{-\infty}^{0} dw^- a^+ (x^+ = 0, x_\perp , w^-) 
  U(x_\perp ) 
  \otimes \int_{0}^{+\infty} dz^-  a^+ (y^+ = 0, y_\perp , z^-) U^\dagger (y_\perp ) \\ 
  & \qquad + U(x_\perp ) \int_{0}^{+\infty} dw^- a^+ (x^+ = 0, x_\perp 
  , w^-) 
  \otimes U^\dagger (y_\perp ) \int_{-\infty}^{0} dz^- a^+ (y^+ = 0, y_\perp , z^-) \\ 
  & \qquad + U(x_\perp ) \int_{0}^{+\infty} dw^- a^+ (x^+ = 0, x_\perp 
  , w^-) 
  \otimes \int_{0}^{+\infty} dz^- a^+ (y^+ = 0, y_\perp , z^-) U^\dagger ( y_\perp ) \\ 
  & \qquad - \int_{-\infty}^{0} dw^- dz^- \theta (z^- - w^-) a^+ (x^+ 
  = 0, x_\perp , w^-) a^+ (x^+ = 0, 
  x_\perp , z^-) U(x_\perp ) \otimes U^\dagger (y_\perp ) \\ 
  & \qquad - \int_{-\infty}^{0} dw^- a^+ (x^+ = 0, x_\perp , w^-) 
  U(x_\perp ) 
  \int_{0}^{+\infty}  dz^- a^+ (x^+ = 0, x_\perp , z^-) \otimes U^\dagger (y_\perp ) \\ 
  & \qquad - U(x_\perp ) \int_{0}^{+\infty} dw^- dz^- \theta (z^- - 
  w^-) a^+ (x^+ = 0, x_\perp , 
  w^-) a^+ (x^+ = 0, x_\perp , z^-) \otimes  U^\dagger (y_\perp ) \\ 
  & \qquad - U( x_\perp )\otimes U^\dagger ( y_\perp ) 
  \int_{-\infty}^{0} dw^- dz^- \theta (w^- - z^-) a^+ 
  (y^+ = 0, y_\perp , w^-) a^+ (y^+ = 0, y_\perp , z^-) \\ 
  & \qquad - U(x_\perp ) \otimes \int_{0}^{+\infty} dw^- a^+ (y^+ = 0, 
  y_\perp , w^-) U^\dagger ( 
  y_\perp )  \int_{-\infty}^{0} dz^- a^+ (y^+ = 0, y_\perp , z^-) \\ 
  & \qquad - U(x_\perp ) \otimes \int_{0}^{+\infty} dw^- dz^- \theta 
  (w^- - z^-) a^+ (y^+ = 0, y_\perp , w^-) a^+ (y^+ = 0, y_\perp , 
  z^-) U^\dagger ( y_\perp ) \big\rangle\, . 
\end{split}  
\label{expansiona} 
\end{equation}

For calculational purposes it is fruitful to separate the different 
contributions into the set terms where the glue interacts with the 
background field (the first four terms in Eq.~(\ref{expansiona}) and 
the set of contributions where there is no such interaction (the other 
terms in Eq.~(\ref{expansiona})). 
 
Let us first consider the contribution where the glue is exchanged 
between the quark in the negative half plane ($x^-<0$) in the 
amplitude and an antiquark in the positive halfplane in the complex 
conjugate amplitude: 
\begin{equation} 
  \int_{-\infty}^{0} dw^- \int_{0}^{+\infty} dz^- \big\langle a_a^+ (x^+ = 0, x_\perp , w^-) a_b^+ (y^+ = 0, y_\perp , 
  z^-) \big\rangle \cdot \big[ ( t_a U(x_\perp)) \otimes ( t_b U^\dagger (y_\perp )) \big]\, . 
\end{equation} 
One notices that only the $\theta(-p^-)$ term in 
Eq.~(\ref{correlatora}) survives and that, Eq.~(\ref{colmat}), the 
color matrix ${\tilde{U}}_{ab}^{-1}(w^-, z^-, z_\perp)$ reduces to 
${\tilde{U}}_{ab} (z_\perp)$. 
\begin{multline} 
  \int_{-\infty}^{0} dw^- \int_{0}^{+\infty} dz^- 
  \big\langle  a_a^+ (x^+ = 0, x_\perp , w^-) 
  a_b^+ (y^+ = 0, y_\perp , z^-) \big\rangle  = \\ 
  = - {\partial}_{x}^{i} {\partial}_{y}^{i} {\int} \frac{dp^-}{p^-} 
  [-\theta (-p^-)] \frac{1}{2(2 \pi)} \frac{1}{(p^-)^2} \int d^2 
  z_{\perp} \int \frac{d^2 p_{\perp}}{(2\pi)^2} e^{+ i p_{\perp} 
    (x_{\perp} - z_{\perp})} 
  \int \frac{d^2 q_{\perp}}{(2\pi)^2} e^{+ i q_{\perp} (z_{\perp} - y_{\perp})} \times \\ 
  \times \Big( \int_{-\infty}^{0} dw^- e^{-i \frac{{p_{\perp}}^2}{2 
      p^-} w^-} \int_{0}^{+\infty} dz^- e^{+i \frac{{q_{\perp}}^2}{2 
      p^-} z^-} \Big) \cdot {\tilde{U}}_{ab} (z_\perp) \, . 
\end{multline} 
The $w^-$ and $z^-$ integrations are easily performed 
\begin{equation} 
  \int_{-\infty}^{0} dw^- e^{-i \frac{{p_{\perp}}^2}{2 p^-} w^-} 
  \int_{0}^{+\infty} dz^- e^{+i \frac{{q_{\perp}}^2}{2 
        p^-} z^-} = - \frac{(2p^-)^2}{{p_{\perp}}^2 {q_{\perp}}^2} \, . 
\end{equation} 
Noting that 
\begin{equation} 
  \int \frac{dp^-}{p^-} \theta (-p^-) = 
  {\int}_{-\infty}^{0} \frac{dp^-}{p^-} = 
  - {\int}_{0}^{+\infty} \frac{dp^-}{p^-} = 
  - \int \frac{dp^-}{p^-} \theta (p^-)\, , 
\label{pminus} 
\end{equation} 
we have 
\begin{equation} 
  \frac{1}{\pi} {\int}_{0}^{+\infty} 
  \frac{dp^-}{p^-} \int d^2 z_{\perp} 
  {\partial}_{x}^{i} {\partial}_{y}^{i} \int 
  \frac{d^2 p_{\perp}}{(2\pi)^2} \frac{1}{{p_{\perp}}^2} 
  e^{+ i p_{\perp} (x_{\perp} - z_{\perp})} \int \frac{d^2 
     q_{\perp}}{(2\pi)^2} \frac{1}{{q_{\perp}}^2} 
   e^{+ i q_{\perp} (z_{\perp} - y_{\perp})} \cdot {\tilde{U}}_{ab} 
  (z_\perp) \, . 
\end{equation} 
The transverse momenta integrals yield 
\begin{equation} 
  \int \frac{d^2 p_{\perp}}{(2\pi)^2} \frac{1}{{p_{\perp}}^2} 
  e^{+ i p_{\perp} (x_{\perp} - z_{\perp})} = - \frac{1}{4 
     \pi} \log (x_\perp - z_\perp)^2 \, . 
\end{equation} 
Taking the derivatives 
\begin{equation} 
   \frac{1}{4 \pi^3} {\int}_{0}^{+\infty} 
  \frac{dp^-}{p^-} \int d^2 z_{\perp} \frac{(x-z)_\perp \cdot 
     (y-z)_\perp}{(x-z)_\perp^2 (y-z)_\perp^2}
   \cdot {\tilde{U}}_{ab} (z_\perp)\, . 
\end{equation} 
We use the following identity valid for $SU(N)$ group 
\begin{align} 
  & {\tilde{U}}_{ab}(z_\perp) (t_a U(x_\perp )) 
  \otimes (t_b U^\dagger (y_\perp )) = \nonumber \\ 
  & \qquad \quad= 2\tr \big[t_a U(z_\perp ) t_b U^\dagger (z_\perp ) 
  \big] 
  (t_a U(x_\perp ))^{\alpha\beta} 
  (t_b U^\dagger (y_\perp ))^{\gamma\delta}  \nonumber \\ 
  & \qquad \quad= \frac{1}{2 N_c} \Big[ N_c {\big( U(z_\perp )\cdot 
    U^\dagger (y_\perp)\big)}^{\alpha\delta} {\big( U^\dagger (z_\perp 
    ) \cdot U(x_\perp )\big)}^{\gamma\beta} - U(x_\perp 
  )^{\alpha\beta} U^\dagger (y_\perp)^{\gamma\delta} \Big] \, . 
\end{align} 

This contribution, the second term in Eq.~(\ref{expansiona}),
therefore is
\begin{multline} 
   \frac{1}{4 \pi^3} {\int}_{0}^{+\infty} \frac{dp^-}{p^-} \int d^2 
  z_{\perp} \frac{(x-z)_\perp \cdot (y-z)_\perp}{(x-z)_\perp^2 
    (y-z)_\perp^2} 
  \times \\ 
  \times \frac{1}{2 N_c} \Big[ N_c {\big( U(z_\perp )\cdot U^\dagger 
    (y_\perp)\big)}^{\alpha\delta} {\big( U^\dagger (z_\perp ) \cdot 
    U(x_\perp )\big)}^{\gamma\beta} - U(x_\perp )^{\alpha\beta} 
  U^\dagger (y_\perp)^{\gamma\delta} \Big]\, . 
\label{exchange1a} 
\end{multline}

The third contribution, due to the exchange of the gluon between the 
quark in the positive half plane and the antiquark in the negative 
halfplane, is calculated similarly with the only difference that now 
${\tilde{U}}_{ab}^{-1}(w^-, z^-, z_\perp) = 
{\tilde{U}}_{ab}^{\dagger}(z_\perp)$ and we pick up the $\theta (p^-)$ 
term in the propagator. 
\begin{align} 
  &\int_{0}^{+\infty} dw^- \int_{-\infty}^{0} dz^- \big\langle a_a^+ 
  (x^+ = 0, x_\perp , w^-) a_b^+ (y^+ = 0, y_\perp , z^-) \big\rangle 
  \cdot 
  \big[(U(x_\perp) t_a) \otimes (U^\dagger (y_\perp ) t_b) \big] = 
  \nonumber \\ 
  &\qquad \qquad = \frac{1}{4 \pi^3} {\int}_{0}^{+\infty} 
  \frac{dp^-}{p^-} \int d^2 z_{\perp} 
  \frac{(x-z)_\perp \cdot (y-z)_\perp}{(x-z)_\perp^2 (y-z)_\perp^2} 
  \times \nonumber \\ 
  &\qquad\qquad\qquad\quad \times \frac{1}{2 N_c} \Big[ N_c {\big( 
    U(x_\perp )\cdot U^\dagger (z_\perp)\big)}^{\alpha\delta} {\big( 
    U^\dagger (y_\perp )\cdot U(z_\perp )\big)}^{\gamma\beta} - 
  U(x_\perp )^{\alpha\beta} U^\dagger (y_\perp)^{\gamma\delta} \Big] 
  \, . 
 \label{exchange2a} 
\end{align}

For the quark self energy correction everything is the same as in 
Eq.~(\ref{exchange1a}) except that the transverse coordinates of the 
two fields coincide ($y_\perp \to {\tilde{x}}_\perp = x_\perp $) 
\begin{align} 
  & - \int_{-\infty}^{0} dw^- \int_{0}^{+\infty} dz^- \big\langle 
  a_a^+ (x^+ = 0, x_\perp , w^-) a_b^+ (y^+ = 0, {\tilde x}_\perp = 
  x_\perp , z^-) \big\rangle \cdot 
  \big[ (t_a U(x_\perp) t_b) \otimes U^\dagger (y_\perp ) \big] = 
  \nonumber \\ 
  & \qquad \qquad = -\frac{1}{4 \pi^3} {\int}_{0}^{+\infty} 
  \frac{dp^-}{p^-} \int d^2 z_{\perp} 
  \frac{1}{(x-z)_\perp^2} \times \nonumber \\ 
  & \qquad\qquad\qquad\quad \times \frac{1}{2 N_c} \Big[ N_c \tr 
  \big((U(x_\perp )\cdot U^\dagger (z_\perp)\big) {\big( U(z_\perp 
    )\big)}^{\alpha\beta} - {\big( U(x_\perp )\big)}^{\alpha\beta} 
  \Big] U^\dagger (y_\perp)^{\gamma\delta} \, . 
\label{nonexchange1a} 
\end{align} 
 
Finally, the antiquark self energy correction
\begin{align} 
  & - \int_{0}^{+\infty} dw^- \int_{-\infty}^{0} dz^- \big\langle 
  a_a^+ (x^+ = 0, y_\perp , w^-) a_b^+ (y^+ = 0, {\tilde y}_\perp = 
  y_\perp , z^-) \big\rangle \cdot 
  \big[ U(x_\perp) \otimes (t_a t_b U^\dagger (y_\perp )) \big] = \nonumber \\ 
  &\qquad \qquad 
  = -\frac{1}{4 \pi^3} {\int}_{-\infty}^{0} 
  \frac{dp^-}{p^-} \int d^2 z_{\perp} 
  \frac{1}{(y-z)_\perp^2} \times \nonumber \\ 
  & \qquad\qquad\qquad\quad \times \frac{1}{2 N_c} \Big[ N_c \tr 
  \big(U^\dagger (y_\perp)\cdot U(z_\perp)\big) {\big( 
    U^\dagger(z_\perp )\big)}^{\gamma\delta} - {\big( U^\dagger 
    (y_\perp )\big)}^{\gamma\delta} \Big] U(x_\perp )^{\alpha\beta} \, 
  . 
\label{nonexchange2a} 
\end{align} 
 
Now let us look at the other set of contributions.  Since for all of 
these there is no interaction with the background we have, 
Eq.~(\ref{colmat}), ${\tilde{U}}_{ab}^{-1}(w^-, z^-, z_\perp) = 
{\delta}_{ab}$. The $z_\perp$ integral is trivial for all terms in 
this set (it yields a $\delta$ function for the transverse momenta). 
However, as we want to combine both sets of contributions in the final 
result, this integral will not be performed. 
 
Consider the first term in Eq.~(\ref{expansiona}) --- the exchange 
between the quark and the antiquark in the negative half-plane --- 
\begin{equation} 
  \int_{-\infty}^{0} dw^- dz^- 
  \big\langle a_a^+ (x^+ = 0, y_\perp , w^-) 
  a_b^+ (y^+ = 0, {\tilde y}_\perp = y_\perp , 
  z^-) \big\rangle \cdot \big[ ( t_a U(x_\perp)) 
  \otimes ( U^\dagger (y_\perp ) t_b) \big]\, . 
\end{equation}

In the correlator part now both $\theta$ functions survive, 
corresponding to the two possible orderings of $w^-$ and $z^-$ 
\begin{multline} 
  \int_{-\infty}^{0} dw^- dz^- 
  \big\langle  a_a^+ (x^+ = 0, x_\perp , w^-) 
  a_b^+ (y^+ = 0, y_\perp , z^-) \big\rangle  = \\ 
  = -\, {\partial}_{x}^{i} {\partial}_{y}^{i} \int \frac{dp^-}{p^-} 
  \frac{1}{2(2 \pi)} \frac{1}{(p^-)^2} \int d^2 z_{\perp} \int 
  \frac{d^2 p_{\perp}}{(2\pi)^2} e^{+ i p_{\perp} (x_{\perp} - 
    z_{\perp})} 
  \int \frac{d^2 q_{\perp}}{(2\pi)^2} 
  e^{+ i q_{\perp} (z_{\perp} - y_{\perp})} \times \\ 
  \times \Big( \int_{-\infty}^{0} dw^- dz^- \big[\theta(w^- - z^-) 
  \theta(p^-) - \theta(z^- - w^-) \theta(-p^-) \big] e^{-i 
    \frac{p_\perp^2}{2 p^-} w^-} e^{+i \frac{{q_{\perp}}^2}{2 p^-} 
    z^-} \Big) \cdot {\delta}_{ab} \, . 
\end{multline} 
The $w^-$ and $z^-$ integrals now have to be performed with more care. 
Doing this we get 
\begin{multline} 
  -\frac{1}{\pi}{\partial}_{x}^{i} {\partial}_{y}^{i} \int 
  \frac{dp^-}{p^-} \int d^2 z_{\perp} \int \frac{d^2 
    p_{\perp}}{(2\pi)^2} e^{+ i p_{\perp} (x_{\perp} - z_{\perp})} 
  \int \frac{d^2 q_{\perp}}{(2\pi)^2} e^{+ i q_{\perp} (z_{\perp} - y_{\perp})} \times \\ 
  \times \Big[ \theta (p^-) \big( \frac{1}{{p_\perp}^2 {q_\perp}^2 }- 
  \frac{1}{{p_\perp}^2 ({q_\perp}^2 - {p_\perp}^2) } \big) + \theta (- 
  p^-) \big( -\frac{1}{{p_\perp}^2 {q_\perp}^2 }- \frac{1}{{q_\perp}^2 
    ({q_\perp}^2 - {p_\perp}^2) } \big) \Big] \cdot \delta_{ab} \, . 
\end{multline} 
Recalling Eq.~(\ref{pminus}) we obtain for the momentum denominators 
\begin{align} 
  & \int \frac{dp^-}{p^-} \Big[ \theta (p^-) \big( 
  \frac{1}{{p_\perp}^2 {q_\perp}^2 } - \frac{1}{{p_\perp}^2 
    ({q_\perp}^2 - {p_\perp}^2) } \big) + \theta (- p^-) \big( 
  -\frac{1}{{p_\perp}^2 {q_\perp}^2 } - 
  \frac{1}{{q_\perp}^2 ({q_\perp}^2 - {p_\perp}^2) } \big) \Big] = \nonumber \\ 
  & \qquad \qquad = {\int}_{0}^{+\infty} \frac{dp^-}{p^-} \Big[ 2 
  \frac{1}{{p_\perp}^2 {q_\perp}^2 } + \big( \frac{1}{{q_\perp}^2 
    ({q_\perp}^2 - 
    {p_\perp}^2)}- \frac{1}{{p_\perp}^2 ({q_\perp}^2 - {p_\perp}^2)} \big) \nonumber \\ 
  &\qquad \qquad = {\int}_{0}^{+\infty} \frac{dp^-}{p^-} 
  \frac{1}{{p_\perp}^2 {q_\perp}^2 }\, . 
\end{align} 
 
The correlator part is then 
\begin{multline} 
  \int_{-\infty}^{0} dw^- dz^- 
  \big\langle  a_a^+ (x^+ = 0, x_\perp, w^- ) 
  a_b^+ (y^+ = 0, y_\perp ,z^-) \big\rangle  = \\ 
  = -\frac{1}{4 \pi^3} {\int}_{0}^{+\infty} \frac{dp^-}{p^-} \int d^2 
  z_{\perp} \frac{(x-z)_\perp \cdot (y-z)_\perp}{(x-z)_\perp^2 
    (y-z)_\perp^2} \cdot {\delta}_{ab} \, . 
\end{multline}

The color algebra raises no problems 
\begin{align} 
  & {\delta}_{ab} (t_a U(x_\perp )) \otimes (U^\dagger (y_\perp ) t_b) 
  = \\ \nonumber & \qquad = {\delta}_{ab} (t_a U(x_\perp 
  ))^{\alpha\beta} (U^\dagger (y_\perp ) t_b)^{\gamma\delta} \\  
  \nonumber & \qquad = \frac{1}{2 N_c} \Big[ N_c 
  {\delta}^{\alpha\delta} {\big( U^\dagger (y_\perp ) \cdot U(x_\perp 
    )\big)}^{\gamma\beta} - U(x_\perp )^{\alpha\beta} U^\dagger 
  (y_\perp)^{\gamma\delta} \Big] 
\end{align} 
and so the first term in Eq.~(\ref{expansiona}) is 
\begin{align} 
  \frac{1}{4 \pi^3} {\int}_{0}^{+\infty} \frac{dp^-}{p^-} \int d^2 
  z_{\perp} & \frac{(x-z)_\perp \cdot 
    (y-z)_\perp}{(x-z)_\perp^2 (y-z)_\perp^2} \times \nonumber \\ 
  & \qquad\times \frac{1}{2 N_c} \Big[ N_c {\delta}^{\alpha\delta} 
  {\big( U^\dagger (y_\perp ) \cdot U(x_\perp )\big)}^{\gamma\beta} - 
  U(x_\perp )^{\alpha\beta} U^\dagger (y_\perp)^{\gamma\delta} \Big]\, 
  . 
\label{exchange3} 
\end{align}

The quark to antiquark exchange in the positive half-plane gives 
\begin{multline}  
  \int_{0}^{+\infty} dw^- dz^- \big\langle a_a^+ (x^+ = 0, y_\perp , 
  w^-) a_b^+ (y^+ = 0, {\tilde y}_\perp y_\perp , z^-) \big\rangle 
  \cdot \big[ (U(x_\perp) t_a) \otimes (t_b U^\dagger (y_\perp ) ) 
  \big] = 
  \\ 
  = -\frac{1}{4 \pi^3} \int\limits_{0}^{+\infty}\! \frac{dp^-}{p^-} 
  \int d^2 z_{\perp} \frac{(x-z)_\perp \cdot 
    (y-z)_\perp}{(x-z)_\perp^2 (y-z)_\perp^2} \times \\ 
  \times \frac{1}{2 N_c} \Big[ N_c {\delta}^{jk} {\big( U(x_\perp ) 
    \cdot U^\dagger (y_\perp ) \big)}^{\alpha\delta} - U(x_\perp 
  )^{\alpha\beta} U^\dagger (y_\perp)^{\gamma\delta} \Big]\, . 
\label{exchange4} 
\end{multline}

Now we combine the two terms that give corrections to the quark line 
--- the fifth and the seventh terms in Eq.~(\ref{expansiona}).  It is 
easy to see that they have the same color structure and will also 
yield the same transverse structure. 
 
The color algebra is trivial 
\begin{equation} 
  {\delta}_{ab} (t_a t_b U(x_\perp )) \otimes U^\dagger (y_\perp )= \delta_{ab} (t_a t_b U(x_\perp ))^{\alpha\beta} 
  (U^\dagger (y_\perp ))^{\gamma\delta} = \frac{(N_c^2 - 1)}{2 N_c} U(x_\perp )^{\alpha\beta} U^\dagger 
  (y_\perp)^{\gamma\delta}\, . 
\end{equation} 
The fifth term is the same as Eq.~(\ref{exchange3}), but with $y_\perp 
\to {\tilde x_\perp} = x_\perp$ and only the $\theta(-p^-)$ term 
surviving 
\begin{multline} 
  - \int_{-\infty}^{0} dw^- dz^- \theta (z^- - w^-) \big\langle a_a^+ 
  (x^+ = 0, x_\perp , w^-) a_b^+ (x^+ = 0, x_\perp , z^-) \big\rangle 
  \cdot 
  \delta_{ab} (t_a t_b U(x_\perp )) \otimes U^\dagger (y_\perp ) = \\ 
  = -\frac{1}{\pi} \partial_{x}^{i}\partial_{y}^{i} 
  {\int}_{0}^{+\infty} \frac{dp^-}{p^-} \int d^2 z_\perp \int 
  \frac{d^2 p_\perp}{(2\pi)^2} e^{+i p_\perp (x_\perp - z_\perp)} 
  \int \frac{d^2 q_\perp}{(2\pi)^2} e^{+i q_\perp (z_\perp - {\tilde x}_\perp)} \times \\ 
  \times \Big[ \frac{1}{p^2_\perp q^2_\perp} + \frac{1}{q^2_\perp 
    (q^2_\perp - p^2_\perp)} \Big] \frac{(N_c^2 -1)}{2 N_c} U(x_\perp 
  )^{\alpha\beta} U^\dagger (y_\perp)^{\gamma\delta}\, . 
\label{virt31} 
\end{multline} 
And the seventh term is the same as Eq.~(\ref{exchange4}) 
\begin{multline}  
  -\int_{0}^{+\infty} dw^- dz^- \theta (z^- - w^-) \big\langle a_a^+ 
  (x^+ = 0, x_\perp , w^-) a_b^+ (x^+ = 0, x_\perp , z^-) \big\rangle 
  \cdot 
  \delta_{ab} (U(x_\perp ) t_a t_b) \otimes U^\dagger (y_\perp ) = \\ 
  = -\frac{1}{\pi} \partial_{x}^{i}\partial_{y}^{i} 
  {\int}_{0}^{+\infty} \frac{dp^-}{p^-} \int d^2 z_\perp \int 
  \frac{d^2 p_\perp}{(2\pi)^2} e^{+i p_\perp (x_\perp - z_\perp)} 
  \int \frac{d^2 q_\perp}{(2\pi)^2} e^{+i q_\perp (z_\perp - {\tilde x}_\perp)} \times \\ 
  \times \Big[\frac{1}{p^2_\perp q^2_\perp} - \frac{1}{p^2_\perp 
    (q^2_\perp - p^2_\perp)} \Big] \frac{(N_c^2 -1)}{2 N_c} U(x_\perp 
  )^{\alpha\beta} U^\dagger (y_\perp)^{\gamma\delta}\, . 
\label{virt32} 
\end{multline} 
 
Adding the two terms Eqs.~(\ref{virt31},\ref{virt32}) and the 
performing the transverse integrations we get 
\begin{equation} 
   \frac{1}{4 \pi^3} {\int}_{0}^{+\infty} 
   \frac{dp^-}{p^-} \int d^2 z_\perp \frac{1}{(x - z)^2_\perp } 
   \frac{(N_c^2 -1)}{2 N_c} 
   U(x_\perp )^{\alpha\beta} U^\dagger (y_\perp)^{\gamma\delta}\, . 
\label{virt3} 
\end{equation} 
 
The correction to the antiquark line --- the eighth and the tenth 
terms in Eq.~(\ref{expansiona}) give similarly 
\begin{equation} 
   \frac{1}{4 \pi^3} {\int}_{0}^{-\infty} 
   \frac{dp^-}{p^-} \int d^2 z_\perp \frac{1}{(y - z)^2_\perp} 
   \frac{(N_c^2 -1)}{2 N_c} 
   U(x_\perp )^{\alpha\beta} U^\dagger (y_\perp)^{\gamma\delta}\, . 
\label{virt4} 
\end{equation} 
 
Finally, combining all the terms together we get 
\begin{align} 
  & {\big\langle V(x_\perp )^{\alpha\beta} V^\dagger (y_\perp 
    )^{\gamma\delta} \big\rangle } - 
  {\big\langle U(x_\perp )^{\alpha\beta} 
    U^\dagger (y_\perp )^{\gamma\delta} \big\rangle } = \nonumber \\ 
  & \qquad =  
  \frac{1}{8 \pi^3} \log (\frac{x_0}{x}) 
  \int d^2 z_\perp \cdot \nonumber \\ 
  & \qquad \qquad \cdot \Bigg\{ \Big[ {\big( U(z_\perp )\cdot 
    U^\dagger (y_\perp)\big)}^{\alpha\delta} {\big( U^\dagger (z_\perp 
    ) \cdot U(x_\perp )\big)}^{\gamma\beta} + {\big( U(x_\perp )\cdot 
    U^\dagger (z_\perp)\big)}^{\alpha\delta} 
  {\big( U^\dagger (y_\perp ) \cdot U(z_\perp )\big)}^{\gamma\beta} \nonumber \\ 
  &\qquad \qquad \qquad - {\delta}^{\alpha\delta} {\big( U^\dagger 
    (y_\perp ) \cdot U(x_\perp )\big)}^{\gamma\beta} - {\big( 
    U(x_\perp ) \cdot U^\dagger (y_\perp ) \big)}^{\alpha\delta} 
  {\delta}^{\beta\gamma} \Big]\cdot 
  \frac{(x-z)_\perp \cdot (y-z)_\perp}{(x-z)_\perp^2 (y-z)_\perp^2} \nonumber \\ 
  & \qquad \qquad \qquad - \Big[ \tr \big( U(x_\perp )\cdot U^\dagger 
  (z_\perp)\big) {\big( U(z_\perp )\big)}^{\alpha\beta} - N_c 
  U(x_\perp )^{\alpha\beta} \Big] U^\dagger 
  (y_\perp)^{\gamma\delta}\cdot 
  \frac{1}{(x - z)^2_\perp} \nonumber \\ 
  & \qquad \qquad \qquad - U(x_\perp)^{\alpha\beta} \Big[ \tr \big( 
  U^\dagger (y_\perp )\cdot U(z_\perp)\big) {\big( U^\dagger (z_\perp 
    )\big)}^{\gamma\delta} - N_c U^\dagger (y_\perp )^{\gamma\delta} 
  \Big] \cdot \frac{1}{(z - y)^2_\perp} \Bigg\} \, . 
\end{align} 
This coincides with the result of Ref.~\cite{Balitskii:1996ub}. When
  comparing this evolution equation with the results of
  Ref.~\cite{Balitskii:1996ub}, one should keep in mind that there the
  evolution is considered with respect to the variable $\zeta$. The
  relation between the two evolution equations is given by
  $\frac{d}{d\ln\frac{1}{x}}\langle\dots\rangle = -2
  \frac{d}{d\ln\zeta}\langle\dots\rangle$.
 
Now taking trace over the color indices we obtain the evolution 
equation for the scattering cross section given in Sec.~\ref{sect:bk}.

\providecommand{\href}[2]{#2}\begingroup\raggedright\endgroup

\end{document}